\documentclass[onecolumn,amsmath,showpacs,11pt]{revtex4}
\usepackage{mathrsfs} 
\usepackage{graphicx}
\usepackage{bm}
\usepackage{float}
\usepackage{fancyhdr}
\usepackage{longtable}
\usepackage{hyperref}
\usepackage{subfigure}
\usepackage{epsfig} 
\usepackage{color,soul}
\usepackage{xcolor}
\definecolor{lb}{RGB}{44, 139, 183}

\begin{document}
	%\preprint{APS/123-QED}
	%%%%%%%%%%%%%%%%%%%%%%%%
	\newcommand{\hs}{\hspace*{0.5cm}}
	\newcommand{\vs}{\vspace*{0.5cm}}
	\newcommand{\be}{\begin{equation}}
		\newcommand{\ee}{\end{equation}}
	\newcommand{\bea}{\begin{eqnarray}}
		\newcommand{\eea}{\end{eqnarray}}
	\newcommand{\ben}{\begin{enumerate}}
		\newcommand{\een}{\end{enumerate}}
	\newcommand{\bde}{\begin{widetext}}
		\newcommand{\ede}{\end{widetext}}
	\newcommand{\nn}{\nonumber}
	\newcommand{\crn}{\nonumber \\}
	\newcommand{\Tr}{\mathrm{Tr}}
	\newcommand{\non}{\nonumber}
	\newcommand{\noi}{\noindent}
	\newcommand{\al}{\alpha}
	\newcommand{\la}{\lambda}
	\newcommand{\bet}{\beta}
	\newcommand{\ga}{\gamma}
	\newcommand{\va}{\varphi}
	\newcommand{\om}{\omega}
	\newcommand{\pa}{\partial}
	\newcommand{\+}{\dagger}
	\newcommand{\fr}{\frac}
	\newcommand{\bc}{\begin{center}}
		\newcommand{\ec}{\end{center}}
	\newcommand{\Ga}{\Gamma}
	\newcommand{\de}{\delta}
	\newcommand{\De}{\Delta}
	\newcommand{\ep}{\epsilon}
	\newcommand{\varep}{\varepsilon}
	\newcommand{\ka}{\kappa}
	\newcommand{\La}{\Lambda}
	\newcommand{\si}{\sigma}
	\newcommand{\Si}{\Sigma}
	\newcommand{\ta}{\tau}
	\newcommand{\up}{\upsilon}
	\newcommand{\Up}{\Upsilon}
	\newcommand{\ze}{\zeta}
	\newcommand{\ps}{\psi}
	\newcommand{\Ps}{\Psi}
	\newcommand{\ph}{\phi}
	\newcommand{\vph}{\varphi}
	\newcommand{\Ph}{\Phi}
	\newcommand{\Om}{\Omega}
	
	%%%%%%%%%%%%%%%%%%%%%%%%
	
	\title{ Physical constraints derived from FCNC in the 3-3-1-1 model}
	\author{N. T. Duy $^{a,b}$}
	\author{Takeo Inami $^{a,c}$}
	\author{D. T. Huong$^{a}$}
	\email{dthuong@iop.vast.ac.vn}
	\email{ntdem@iop.vast.ac.vn}
	
	\affiliation{
		$^a$ Institute of Physics, VAST, 10 Dao Tan, Ba Dinh, Hanoi, Vietnam\\
		$^b$ Graduate University of Science and Technology,
		Vietnam Academy of Science and Technology,
		18 Hoang Quoc Viet, Cau Giay, Hanoi, Vietnam \\
		$^c$ Theoretical Research Division, Nishina Center, RIKEN, Wako 351-0198, Japan}
	\date{\today}

	\date{\today}
	
	\begin{abstract}
	We investigate several phenomena related to FCNCs in the $\text{3-3-1-1}$ model. The sources of FCNCs at the tree-level from both the gauge and Higgs sectors are clarified. Experiments on the oscillation of mesons most stringently constrain the tree-level FCNCs. The lower bound on the new physics scale is imposed more tightly than in the previous, $\text{M}_{\text{new}}>12 $ \text{TeV}.  Under this bound, the tree-level FCNCs make a negligible contribution to the $\text{Br}(B_s \rightarrow \mu^+ \mu^-)$, $\text{Br}(B \rightarrow K^{*} \mu^+ \mu^-)$ and $\text{Br}(B^{+}\rightarrow K^{+}\mu^{+}\mu^{-})$. The branching ratio of radiative decay $b \rightarrow s \gamma$ is enhanced by the ratio $\frac{v}{u}$ via diagrams with the charged Higgs mediation. In contrast, the charged currents of new gauge bosons significantly contribute to the decay process $\mu \rightarrow e \gamma$. 
	\end{abstract}
	
	\maketitle
	\section{\label{intro}Introduction}
	The analysis of phenomena related to flavor-changing neutral currents (FCNCs) plays an important role in constraining the parameters of the Standard Model (SM) and testing physics beyond the standard model (BSM). In recent years, the most extensively studied processes related to FCNCs in B-physics, particularly the exclusive $b \rightarrow s$ transition. The first place to look for new physics (NP) in $b \rightarrow s$ transitions is $B_q-\bar{B}_q$ mixing with $q=d,s$. The mass splitting $\Delta M_d$ has been measured with high precision \cite{Bdmixing}, whereas the measurement of $\Delta M_s$ \cite{Bsmixing,Bsmixing1} is complicated because of the rapid oscillation of ${B}_s$ meson. 
	%The observed $\Delta M_q$ agrees well with the SM predictions. 
	The measurement results of Br$(B_s \rightarrow \mu^+ \mu^-)$ \cite{Bsmm1,LHCb2, LHCb3, LHCb2021}, Br$(b \rightarrow s \gamma)$ \cite{bsgamma,bsgamma1, bsgamma2,bsgamma4}, are almost in agreement with the SM predictions. However, some small tensions related to the above processes have been persisted  and confirmed by independent measurements. These tensions can be understood due to uncertainties of the form factors, CKM elements, or by the presence of NP.  Moreover, the ratios of branching fractions $R_{K}, R_{K^*}$, and several observables of the $ B \rightarrow K(K^*) l^+l^-$ ($l=\mu,e$) decays have been determined  \cite{AaiJ:2017vbb,Abdesselam:2019wac, AaiJ:2021, Aaij:2015dea, Khachatryan:2015isa, Wehle:2016yoi, Sirunyan:2017dhj, Aaboud:2018krd, Aaij:2020nrf, bsll2,bsll2-bs1,bsll2-bs2,bsll2-bs3}. All the results of these measurements have confirmed the deviation from the predictions of the SM. Unlike the angular observables, the various ratios of branching fractions can not be explained via underestimating hadron effects. This result has inspired physicists to investigate these decay processes and see whether some NP models can better explain the experimental data.

	%A common feature of all popular weakly-coupled extensions of the SM is an enlarged Higgs sector and extended gauge symmetry.
	Recently, P.V. Dong and his collaborators have pointed out the simple extension of the SM in which the gauge symmetry has been extended to the $SU(3)_C \times SU(3)_L \times U(1)_X \times U(1)_N$ group, referred to as the $ \text{3-3-1-1}$ model. This model contains both mathematical and phenomenological aspects of the 3-3-1 model \cite{331a,331b,331c,331d,331e,331f}. Therefore, the $\text{3-3-1-1}$ model has all the good features of the $\text{3-3-1}$ models \cite{3311,3311a,3311b,3311c}. The difference between the $\text{3-3-1-1}$ model and previous $\text{3-3-1}$ versions is the nature of $B-L$ symmetry . In the \text{3-3-1-1} model, the $B-L$ symmetry is known as a non-commutative gauge symmetry. Therefore, there exists a unification between the electroweak and $B-L$ interactions \cite{3311d}, which is similar to the Glashow-Weinberg-Salam theory. In addition, the model also provides a natural, comprehensive scenario to account for neutrino masses, dark matter, inflation, and leptogenesis \cite{3311d}. 
	
	Another feature of the 3-3-1-1 model is that flavor-violating interactions appear in both the quark and lepton sectors. The quark families transform differently under $SU(3)_L$. So, they lead to tree-level flavor-changing neutral currents (FCNCs) that couple to the new neutral gauge bosons, $Z_2, Z_N$, and the new neutral Higgs bosons. The role of FCNCs coupled to $Z_2, Z_N$ in the oscillation of mesons has been studied in \cite{3311a}, \cite{3311f}. The authors only focused on the NP short-distance tree-level contribution caused by new neutral gauge bosons to the mass difference of mesons in those studies. The authors used only the NP contributions to compare with the experimental values. Thus, they have pointed out the lower bound on the NP scale in the TeVs. However, considering all NP and SM contributions to the meson oscillations, the lower bound may be more constrained than the previously known ones \cite{3311a}, \cite{3311f}.
	
	% which often puts a severe constraint on the BSM and the $\mu \rightarrow e \gamma$ decay.	 

	In this paper, we study all tree-level FCNCs associated with both Higgs and gauge bosons. The contributions coming from the  FCNCs combined with these of SM are subject to strong constraints from meson mixing parameters. Phenomenological aspects related to FCNCs at tree-level, namely $B_s \rightarrow \mu^{+} \mu^{-}$, $B \to K^{*} \mu^+ \mu^-$ and $B^{+}\to K^{+}\mu^{+}\mu^{-}$ decays are expensive goals. Additionally, the 3-3-1-1 model predicts the existence of new charged particles, such as new non-Hermitian gauge bosons $Y_{\mu}^{\pm}$,  the charged Higgs bosons $H_{4,5}^{\pm}$. They couple to both SM quarks, leptons to new heavy quarks, leptons, respectively. These interactions are the source for yielding the charged lepton flavor violation (LFV) processes $l_i \rightarrow l_j \gamma$
	and $b \rightarrow s \gamma$ decay.
	
	% Based on these studies, we can show that the large FCNC can be avoided even considering all contributions mentioned above. 
	
	We organize our paper as follows. In Sec. \ref{3311}, we briefly overview the $\text{3-3-1-1}$ model. In Sec \ref{FCNC-treelevel}, we describe the tree-level FCNCs and study their effects on the mass difference of mesons. We predict the NP contributions to the rare decays of $B_s \rightarrow \mu^{+} \mu^{-} $, $B \rightarrow K^{*} \mu^+ \mu^-$ and $B^{+}\to K^{+}\mu^{+}\mu^{-}$ processes based on the constrained parameter space.  
	%In this section, we concentrate on studying the mass difference of mesons within the tree-level FCNC, the rare decays $B_s %\rightarrow \mu^{+} \mu^{-} $ and $B \rightarrow K^{(*)} \mu^+ \mu^-$.
	Sec. \ref{Radiative} studies the one-loop calculation of the relevant Feynman diagrams, which relate to the $b\rightarrow s\gamma$ and $\mu \rightarrow e \gamma$. The consequences of the parameters on the branching ratio of these decays are implied from the experimental data studied. Our conclusions are given in Sec.\ref{Conclusion}.

	\section{\label{3311}A summary of the 3-3-1-1 model}
	\subsection{Symmetry and particle content}
	The gauge symmetry of the model is  $SU(3)_C \times SU(3)_L\times U(1)_X \times U(1)_N$, where $SU(3)_C$ is the color group, $SU(3)_L$ is an extension of the  $SU(2)_L$ weak-isospin, and $U(1)_X$, $U(1)_N$ define the electric charge $Q$ 
	and $B-L$ operators \cite{3311f} as follows
	\bea Q=T_3+\beta T_8+ X, \hs B-L=\beta^\prime T_8+N, \eea
	%	\bea
	%	\text{Q}= \text{T}_3+\beta \text{T_8} +\text{X}, \hs  \text{B}-\text{L}= \beta^\prime \text{T}_8+\text{N},
	%	\eea
	where $\beta, \beta^\prime$ are coefficients, and both are free from anomalies. The parameters $\beta,\beta^{\prime}$ determine the $Q$ and $ B-L$ charges of new particles. In this work, we consider the model with $\beta=-\fr{1}{\sqrt{3}}$. This is the simple 3-3-1-1 model for dark matter \cite{3311}. The leptons and quarks, free of all gauge anomalies, transform as
	\bea \psi_{aL} && = (\nu_{aL}, e_{aL}, (N_{aR})^c)^T \sim (1,3, -1/3,-2/3), \hs \nu_{aR}\sim (1,1,0,-1), \hs  e_{aR} \sim (1,1, -1,-1), \crn  Q_{\al L} &&=(d_{\al L},-u_{\al L}, D_{\al L})^T \sim (3,3^*,0,0), \hs Q_{3L}=(u_{3L},d_{3L},U_L)^T\sim (3,3,1/3,2/3), \crn u_{aR}&& \sim (3,1,2/3,1/3),\hs
	d_{aR}\sim (3,1,-1/3,1/3), \hs U_R \sim (3,1,2/3,4/3),\hs D_{aR}\sim (3,1,-1/3,-2/3), \crn \eea
	where $a=1,2,3$,  $\al=1,2$ are the generation indexes. The scalar sector, which is necessary for realistic symmetry breaking and mass generation, consists of the following Higgs fields \cite{3311} 
	\bea \eta^T && =(\eta^0_1, \eta_2^-,\eta_3^0)^T \sim (1,3,-1/3,1/3), \hs \rho^T=(\rho_1^+,\rho_2^0,\rho_3^+)^T \sim (1,3,2/3,1/3), \crn  \chi^T && =(\chi_1^0, \chi_2^-, \chi_3^0)^T \sim (1,3,-1/3,-2/3),\hs \phi \sim(1,1,0,2). \eea 
	The electrically-neutral scalars can develop vacuum expectation values (VEVs) \bea  <\eta_{1}^0>= \fr{u}{\sqrt{2}}, \hs  <\rho_2^0>=\fr{v}{\sqrt{2}}, \hs  <\chi_3^0> =\fr{w} {\sqrt{2}}, \hs  <\phi>=\frac{\La}{\sqrt{2}},\eea 
	and break the symmetry of model via the following scheme 
	\bc
	\begin{tabular}{c}  
		$SU(3)_C\otimes SU(3)_L\otimes U(1)_{X} \otimes U(1)_N$\\
		$  \downarrow  \La$\\
		$SU(3)_C\otimes SU(3)_L  \otimes U(1)_X  \otimes P$\\
		$\downarrow w $\\ 
		$SU(3)_C\otimes SU(2)_L \otimes U(1)_{B-L}\otimes P$\\
		$\downarrow u,v $\\ 
		$SU(3)_C \otimes U(1)_{Q}\otimes P$, \\ 
	\end{tabular}
	\ec
	where $P$ is understood as the matter parity (W-parity) and takes the form: $P=(-1)^{3(B-L)+2s}$. All SM particles have W-parity of $+1$ (called even \text{W}-particle) while new fermions have W-parity of $-1$ (called odd \text{W}-particle). With W-parity preserved, the lightest odd \text{W}-particle can not decay. If the lightest particle has a neutral charge, it may account for dark matter (see \cite{3311}). The VEVs, $u,v$, break the electroweak symmetry and generate the mass for SM particles with the consistent condition: $u^2+v^2=246^2 \ \text{GeV}^2$. The VEVs, $w,\La,$  break $SU(3)_L, U(1)_N$ groups and generate the mass for new particles. For consistency, we assume $ w,\La \gg u,v $.
	\subsection{Scalar sector}	
	Let us rewrite the scalar potential \cite{3311a}, \cite{3311b} that consists of three terms, $V= V(\phi)+ V(\eta, \rho, \chi) +V_{\text{mix}}$, where
	\bea
	V(\phi) && = \mu_\phi^2 \phi^\dag \phi +\lambda (\phi^\dag \phi)^2, \crn
	V(\eta, \chi, \rho) && = \mu_1^2 \rho^\dag \rho+\mu_2^2 \chi^\dag \chi + \mu^2_3 \eta^\dag \eta +\la_1(\rho^\dag \rho)^2+\la_2(\chi^\dag \chi)^2+ \la_3 (\eta^\dag \eta)^2, \crn
	V_{\text{mix}}&& =\la_4 (\rho^\dag \rho)(\chi^\dag \chi)+\la_5(\rho^\dag \rho)(\eta^\dag \eta)+\la_6 (\chi^\dag \chi)(\eta^\dag \eta) +\la_7(\rho^\dag \chi)(\chi^\dag \rho) +\la_8 (\rho^\dag \eta)(\eta^\dag \rho)\crn &&+ \la_9(\chi^\dag \eta)(\eta^\dag \chi)+\la_{10}(\phi^\dag \phi)(\rho^\dag \rho)+\la_{11}(\phi^\dag \phi)(\chi^\dag \chi)+(f \epsilon^{mnp}\eta_m\rho_n \chi_p +H.c.). 
	\eea
	Due to the \text{W}-parity conservation, only neutral scalar fields carrying \text{W}-parity of $+1$ can develop VEV. After symmetry breaking, there is no mixing between the even and odd \text{W}-fields (see in \cite{3311b} ). For the even \text{W}-particle spectrum, the model has predicted 
	\begin{itemize}
		\item Four neutral physical particles with \text{CP}-even, one identified as the SM-like Higgs boson $H$ and the three remaining particles, $H_i, i=1,2,3$, are new heavy fields, having the following form  
		\bea
		H &&= \fr{u\Re (\eta_1^0)+v \Re (\rho_2^0) }{\sqrt{u^2+v^2}}, \hs \hs \hs  H_1 = \fr{-v\Re (\eta_1^0)+u \Re (\rho_2^0) }{\sqrt{u^2+v^2}}, \crn
		H_2&&= \cos \varphi \Re (\chi_3) +\sin\varphi \Re (\phi), \hs H_3= -\sin\varphi \Re (\chi_3)+\cos \varphi \Re (\phi),
		\eea
		where $\tan (2\varphi)= -\fr{\la_{11}w \La}{\la \La^2 -\la_2 w^2}$.
		\item One neutral CP-odd particle
		\bea
		\mathcal{A}\simeq \frac{v \Im(\eta_1)+u \Im(\rho_2)}{\sqrt{u^2+v^2}}.
		\eea
		\item Two charged fields that are given as follows
		\bea
		H_4^\pm  &&= \fr{v \chi_2^\pm +\omega \rho_3^\pm}{\sqrt{v^2+\omega^2}}, \hs H_5^\pm =\fr{v\eta_2^\pm +u \rho_1^\pm}{\sqrt{u^2+v^2}}. 
		\eea
	\end{itemize}
	For the odd W-particle spectrum, there exists a complex scalar particle
	\bea H^{'0}= \frac{1}{\sqrt{u^2+w^2}}\left(u \chi_1^{0*}+w \eta^0_3 \right).\eea  
	
	For convenience, we list a few mass expressions for the physical fields that we will use for the calculations below
	\bea m_{H_1}^2 && =-\fr{fw}{\sqrt{2}}\left(\fr{v}{u}+\fr{u}{v}\right), \hs \hs \hs \hs \hs m_{\mathcal{A}}^2=-\fr{f}{\sqrt{2}}\left(\fr{uw}{v}+\fr{vw}{u}+\fr{uv}{w}\right),\crn 
	m^2_{H_4}&& =\left(\frac{\la_7}{2}-\frac{fu}{\sqrt{2}vw} \right) \left(v^2+w^2 \right),  \hs m^2_{H_5}=\left( \frac{\la_8}{2}-\frac{fw}{\sqrt{2}uv}\right)\left(u^2+v^2 \right). \label{m-scalar} \eea
	\subsection{Fermion masses}
	The Yukawa interactions in the quark sector are written in \cite{3311} as follows
	\bea
	\mathcal{L}^{\text{quark}}_{\text{Yukawa}}&& = h^U\bar{Q}_{3L}\chi U_R + h^D_{\al \beta}\bar{Q}_{\al L} \chi^* D_{\beta R}+ h^u_a \bar{Q}_{3L}\eta u_{aR} \crn && +h^d_a\bar{Q}_{3L}\rho d_{aR} + h^d_{\al a} \bar{Q}_{\al L}\eta^* d_{aR} +h^u_{\al a } \bar{Q}_{\al L}\rho^* u_{aR} +H.c..\label{yukquark}
	\eea 
	After symmetry breaking, the up-quarks and down-quarks receive mass. Their mixing mass matrices have the following form
	\bea
	m_{\al a}^u=\fr{1}{\sqrt{2}}h_{\al a}^u v, \hs m_{3a}^u =-\fr{1}{\sqrt{2}}h_a^u u, \hs m_{\al a}^d=-\fr{1}{\sqrt{2}}h_{\al a}^d u, \hs m_{3a}^d =-\fr{1}{\sqrt{2}}h_a^d v. 
	\label{massq1}\eea
	In the general case, these matrices are not flavor-diagonal. They can be diagonalized by the unitary  matrices $V_{u_{L,R}}, V_{d_{L,R}}$ as 
	\bea
	V_{u_L}^\dag m^u V_{u_R}=\mathcal{M}_u =\text{Diag}(m_{u_1}, m_{u_2},m_{u_3}), \hs V_{d_L}^\dag m^d V_{d_R}=\mathcal{M}_d =\text{Diag}(m_{d_1}, m_{d_2},m_{d_3}).
	\label{massq2}\eea
	It means that the mass eigenstates relate to the flavor states by
	\bea
	u^\prime_{L,R}&& =(u_{1L,R}^\prime, u_{2L,R}^\prime, u_{3L,R}^\prime)^T= V^{\dagger}_{u_{L,R}}(u_{1L,R},u_{2L,R},u_{3L,R})^T,\crn  d^\prime_{L,R}&& =(d_{1L,R}^\prime, d_{2L,R}^\prime, d_{3L,R}^\prime)^T=V^{\dagger}_{d_{L,R}}(d_{1L,R},d_{2L,R},d_{3L,R})^T.
	\label{massq3}\eea
	The CKM matrix is defined as $V_{\text{CKM}}=V_{u_{L}}^\dag V_{d_L}$.
	
	The Yukawa interactions for leptons are written by
	\bea \mathcal{L}_{\mathrm{Yukawa}}^{\text{lepton}}&=&h^e_{ab}\bar{\psi}_{aL}\rho e_{bR} +h^\nu_{ab}\bar{\psi}_{aL}\eta\nu_{bR}+h'^\nu_{ab}\bar{\nu}^c_{aR}\nu_{bR}\phi +H.c..\label{yuklep1}
	\eea 
	The charged leptons have a Dirac mass  
	$[M_\text{l}]_{ab}=-\fr{h_{ab}^e v}{\sqrt{2}}$. The flavor states $e_a$ are related to the physical states $e_a^\prime$ by using two unitary matrices $U^l_{L,R}$ as
	\bea
	e_{aL}=(U^l_L)_{ab}e^\prime_{bL}, \hs e_{aR}=(U^l_R)_{ab}e^\prime_{bR}.
	\eea
	The neutrinos have both Dirac and Majorana mass terms. In the flavor states, $n_L=(\nu_L, \nu_R^c)^T$, the neutrino mass terms can be written as follows
	\bea
	\mathcal{L}^\nu_{\text{mass}}=-\fr{1}{2} \bar{n}_L \left(%
	\begin{array}{cc}
		0 & M^D_\nu\\
		(M_\nu^{D})^T & M_R^\nu 
	\end{array}% 
	\right)n_L+H.c.=-\fr{1}{2} \bar{n}_L M^\nu n_L +H.c.,
	\eea
	where $[M^D_\nu]_{ab}= -\fr{h_{ab}^\nu}{\sqrt{2}}u $, $[M_{\nu}^R]_{ab}=-\sqrt{2}h^{\prime \nu}_{ab} \La \label{Ma1}$. 
	The mass eigenstates $n_L^\prime$ are related to the neutrino flavor states as $n_L^\prime = U^{\nu\dag} n_L$, where $U^\nu$ is a $6 \times 6$ matrix and written in terms of
	\bea
	U^\nu =\left(%
	\begin{array}{cc}
		U^\nu_L & V^\nu\\
		(V^{\nu})^ T & U_R^\nu 
	\end{array}% 
	\right).
	\eea 
	The new neutral fermions $N_{a}$ are a Majorana field, and they obtain their mass via effective interactions \cite{3311a, 3311b}. We suppose that the flavor states $N_a$ relate to the mass eigenstates $N_a^\prime$ by using the unitary matrices $U^N_{L,R}$ as 
	\bea
	N_{aL}=(U_L^N)_{ab}N^\prime_{bL}, \hs N_{aR}=(U^N_R)_{ab}N^\prime_{bR}.
	\eea 
	\subsection{Gauge bosons}
	Let us review the characteristics of the gauge sector. In addition to the SM gauge bosons, the 3-3-1-1 model also predicts six new gauge bosons: $X^{0,0*}, Y^\pm, Z_2, Z_N$. The gauge bosons are even W-parity except for the $X, Y$ gauge bosons that carry odd W-parity. The masses of new gauge bosons have been given in \cite{3311a}, \cite{3311b} as
	\bea 
	m_{Z_2}^2 &\simeq& \fr{g^2}{18}\left\{(3+t_X^2)w^2+4t_N^2(w^2+9\La^2)  \right.  \nonumber \\  
	&&-\left. \sqrt{[(3+t_X^2)w^2-4t_N^2(w^2+9\La^2)]^2+16(3+t_X^2)t_N^2w^4}\right\}, \\
	m_{Z_N}^2 &\simeq& \fr{g^2}{18}\left\{(3+t_X^2)w^2+4t_N^2(w^2+9\La^2)  \right.  \nonumber \\  
	&&+\left. \sqrt{[(3+t_X^2)w^2-4t_N^2(w^2+9\La^2)]^2+16(3+t_X^2)t_N^2w^4}\right\},
	\nonumber \\ 	m_W^2 &=&\fr{g^2}{4}(u^2+v^2), \hs  m^2_X =\frac{g^2}{4}\left (u^2+w^2 \right), \hs m^2_Y =\frac{g^2}{4}\left(v^2+w^2\right).
	\eea  
	\section{\label{FCNC-treelevel} Rare processes mediated by new gauge bosons and new scalars at the tree-level}
	\subsection{Meson mixing at tree level \label{BBmix}}
	%The \text{FCNCs} coupling to the new neutral gauge boson $Z_2$ and $Z_N$ has been studied in \cite{3311a},\cite{3311f}.
    In previous works \cite{3311a},\cite{3311f}, the authors have considered the FCNCs that couple to the new neutral gauge bosons $Z_2$ and $Z_N$ at tree-level. Due to the different arrangements between generations of quarks, the SM quarks couple to two Higgs triplets. Therefore, there exist FCNCs coupled to the new neutral Higgs bosons at tree-level. These interactions derive from the Yukawa Lagrangian (\ref{yukquark}). After rotating to the physical basis via using Eqs. (\ref{massq1}),(\ref{massq2}), (\ref{massq3}), we obtain the following 
	%off-diagonal couplings are generally produced, inducing FCNCs at tree-level as follows
	\bea
	\mathcal{L}_{\text{NC}}^{\text{Higgs}}&&=-\fr{g}{2m_W}\left(\bar{d}^{\prime}_L\mathcal{M}_d d_R^\prime+ \bar{u}^{\prime}_L \mathcal{M}_u u_R^\prime \right)H +\fr{g}{2 m_W}\left( t_\beta \bar{d}^{\prime}_L\mathcal{M}_d d_R^\prime-\fr{1}{t_\beta}\bar{u}^{\prime}_L \mathcal{M}_u u_R^\prime \right)H_1 \crn && +\fr{ig}{2 m_W}\left( t_\beta \bar{d}^{\prime}_L\mathcal{M}_d d_R^\prime+\fr{1}{t_\beta}\bar{u}^{\prime}_L \mathcal{M}_u u_R^\prime \right)\mathcal{A}+\fr{g}{2m_W}\left(\bar{d}^{\prime}_L\Ga^d d^\prime_R+\bar{u}^{\prime}_L\Ga^u u^\prime_R\right)H_1\crn && +\fr{ig}{2m_W}\left(\bar{d}^{\prime}_L\Ga^d d^\prime_R-\bar{u}^{\prime}_L\Ga^u u^\prime_R\right)\mathcal{A}+ H.c.,
	\label{quark1}\eea 
	where $t_\beta= \tan \beta =\fr{v}{u}$, and $\Ga^u, \Ga^d$ are defined as: 
	\bea
	\Ga^u_{ ij}&& =\fr{2}{s_{2\beta}}(V_{u_L}^\dag)_{i 3} (V_{u_L})_{3k} m_{u_k}(V_{u_R}^\dag)_{ka}(V_{u_R})_{a j},\crn
	\Ga^d_{ij}&& =-\fr{2}{s_{2\beta}}(V_{d_L}^\dag)_{i 3}(V_{d_L})_{3k}m_{d_k}(V_{d_R}^\dag)_{ka}(V_{d_R})_{a j}.
	\eea
	The first three terms of Eq. (\ref{quark1}) are proportional to the quark mass matrices, and thus they are flavor-conserving interactions. The remaining terms are the FCNCs coupled to the new neutral Higgs bosons, including CP-even $H_1$ and CP-odd $\mathcal{A}$. 
	%off-diagonal couplings are generally produced, inducing FCNCs at tree-level as follows  
	%The flavor-violating interactions also are produced by the new gauge bosons $Z_2, Z_{N}$.
	
	%off-diagonal couplings are generally produced, inducing FCNCs at tree-level as follows
	The Lagrangian of tree-level FCNCs mediated by $Z_2, Z_N$, which has been studied in \cite{3311a}, has the following form 
	\bea
	\mathcal{L}_{\text{FCNC}}^{\text{gauge}}&=& -\sum_{q^\prime=u^\prime,d^\prime}\Theta_{ij}^q \left\{\bar{q}^{\prime}_{iL}\ga^\mu q^{\prime}_{jL}(g_2 Z_{2 \mu}+g_N Z_{N \mu}) \right\},
	\label{quark2a}\eea
	where \bea \Theta_{ij}^q&& =  \frac{1}{\sqrt{3}}(V_{q_L}^*)_{3i}(V_{q_L})_{3j}, \hs  g_2=g\left(\cos\xi \fr{1}{\sqrt{1-t_w^2/3}}+\sin\xi \fr{2t_N}{\sqrt{3}} \right), \crn g_N &&=g\left(-\sin\xi \fr{1}{\sqrt{1-t_w^2/3}}+\cos\xi \fr{2t_N}{\sqrt{3}} \right).\eea
	$\xi$ is a mixing angle that is determined by $\tan 2\xi=\frac{4 \sqrt{3+t_X^2}t_N w^2}{(3+t_X^2)w^2-4t_N^2(w^2+9\La^2)}$, $t_N=\fr{g_N}{g}$, and $t_X=\fr{g_X}{g}=\fr{\sqrt{3}s_W}{\sqrt{3-4s_W^2}}$ with $s_W=\sin \theta_W$. 
	
	We now investigate the impact of FCNCs associated with both new gauge and scalar bosons on the oscillation of mesons. From FCNCs given in Eqs. (\ref{quark1})-(\ref{quark2a}), we obtain the effective Lagrangian that affects the meson mixing as 
	%The contributions of FCNC Lagrangian (\ref{quark2a}) to the mass difference of the mesons systems %$K^0-\bar{K^0}$,$B_s^0-\bar{B}_s^0$ and $B_d^0-\bar{B}_d^0$ have been studied in \cite{3311a}, \cite{3311f}. However, the %flavor-violating interactions given in (\ref{quark1}) also give the new contributions to the mass difference of the mesons. %It is necessary to take account of the two contributions in a single analysis. It is straightforward to derive the %effective Lagrangian for the meson mixing from Eqs (\ref{quark1})-(\ref{quark2a}), %namely                                                                                                                                                              
	\bea
	\mathcal{L}_{\text{effective}} && = \fr{g^2}{4m_W^2}\left\{(\Gamma^q_{ij})^2 \left (\frac{1}{m_{H_1}^2}- \frac{1}{m_{\mathcal{A}}^2} \right)\left(\bar{q}^{\prime}_{iL} q^\prime_{jR} \right )^2+(\Gamma_{ji}^{q  *})^2 \left (\frac{1}{m_{H_1}^2}-\frac{1}{m_{\mathcal{A}}^2}\right)\left (\bar{q}^{\prime}_{iR} q^\prime_{jL} \right)^2\right\}\crn &&+ \fr{g^2}{4m_W^2}\left\{\Gamma_{ji}^{q*}\Gamma_{ij}^{q} \left(\fr{1}{m_{H_1}^2}+\fr{1}{m_{\mathcal{A}}^2} \right)(\bar{q}^{\prime}_{iL}q^\prime_{jR})(\bar{q}^{\prime}_{iR} q^\prime_{jL})+\Gamma_{ji}^{q *}\Gamma_{ij}^q\left(\fr{1}{m_{H_1}^2}+\fr{1}{m_{\mathcal{A}}^2} \right)(\bar{q}^{\prime}_{iR} q^\prime_{jL})(\bar{q}^{\prime}_{iL} q^\prime_{jR})\right\} \crn && -\Theta_{ij}^2\left(\fr{g_2^2}{m_{Z_2}^2}+\fr{g_N^2}{m_{Z_N}^2} \right)(\bar{q}^{\prime}_{iL} \ga^\mu q^\prime_{jL})^2 , 
	\label{dmass1}\eea
	with $q$ denoting either $u$ or $d$ quark. This Lagrangian gives contributions to the mass difference of the meson systems as given 
	\bea
	(\Delta m_K)_{\text{NP}} &&=\Re \left\{ \fr{2}{3}\Theta_{12}^2 \left(\fr{g_2^2}{m_{Z_2}^2}+\frac{g_N^2}{m_{Z_N}^2} \right)+\fr{5g^2}{48 m_W^2}\left((\Gamma_{12}^{d})^2+(\Gamma_{21}^{d*})^2\right)\left(\fr{1}{m_{H_1}^2}-\fr{1}{m_{\mathcal{A}}^2} \right) \left(\fr{m_K}{m_s+m_d} \right)^2 \right\}m_K f_K^2\crn && -\Re\left\{\frac{g^2\Gamma_{21}^{d*}\Gamma_{12}^d}{4m_W^2}\left(\fr{1}{m_{H_1}^2}+\fr{1}{m_{\mathcal{A}}^2} \right)\left(\fr{1}{6}+\fr{m_K^2}{(m_s+m_d)^2} \right) \right\}m_K f_K^2, \crn
	(\Delta m_{B_d})_{\text{NP}} &&=\Re \left\{ \fr{2}{3}\Theta_{13}^2 \left(\fr{g_2^2}{m_{Z_2}^2}+\frac{g_N^2}{m_{Z_N}^2} \right)+\fr{5g^2}{48 m_W^2}\left((\Gamma_{13}^{d})^2+(\Gamma_{31}^{d*})^2 \right)\left(\fr{1}{m_{H_1}^2} -\fr{1}{m_{\mathcal{A}}^2}\right)\left(\fr{m_{B_d}}{m_b+m_d} \right)^2 \right\}m_{B_d} f_{B_d}^2\crn && -\Re\left\{\frac{g^2\Gamma_{31}^{d*}\Gamma_{13}^d}{4m_W^2}\left(\fr{1}{m_{H_1}^2}+\fr{1}{m_{\mathcal{A}}^2} \right)\left(\fr{1}{6}+\fr{m_{B_d}^2}{(m_b+m_d)^2} \right) \right\}m_{B_d} f_{B_d}^2,\crn
	(\Delta m_{B_s})_{\text{NP}} &&=\Re \left\{ \fr{2}{3}\Theta_{23}^2 \left(\fr{g_2^2}{m_{Z_2}^2}+\frac{g_N^2}{m_{Z_N}^2} \right)+\fr{5g^2}{48 m_W^2}\left((\Gamma_{32}^{d*})^2 +(\Gamma_{23}^{d})^2\right)\left(\fr{1}{m_{H_1}^2}-\fr{1}{m_{\mathcal{A}}^2} \right)\left(\fr{m_{B_s}}{m_s+m_b} \right)^2 \right\}m_{B_s} f_{B_s}^2\crn && -\Re\left\{\frac{g^2\Gamma_{32}^{d*}\Gamma_{23}^d}{4m_W^2}\left(\fr{1}{m_{H_1}^2}+\fr{1}{m_{\mathcal{A}}^2} \right)\left(\fr{1}{6}+\fr{m_{B_s}^2}{(m_s+m_b)^2} \right) \right\}m_{B_s} f_{B_s}^2.
	\label{d2mass}\eea
	We would like to remind the reader that the theoretical predictions of the meson mass differences account for both SM and all tree-level contributions. It hints that meson mass differences can be separated as 
	\be \Delta m_{K,B_d,B_s}=(\Delta m_{K,B_d,B_s})_{\text{SM}}+(\Delta m_{K,B_d,B_s})_{\text{NP}}, \label{total}\ee 
	where the \text{SM} contributions to the meson mass differences are given by \cite{Meson1a},\cite{Meson1b} 
	\be (\Delta m_K)_{\mathrm{SM}}=0.467\times 10^{-2}/ps,\hs (\Delta m_{B_d})_{\mathrm{SM}}=(0.575^{+0.093}_{-0.090})/ps,\hs (\Delta m_{B_s})_{\mathrm{SM}}=(18.6^{+2.4}_{-2.3})/ps.\ee 
	The theoretical predictions, given in Eq. (\ref{total}), are compared with the experimental values as  given in \cite{HFLAV},\cite{pdg}
	\bea 
	(\Delta m_K)_{\mathrm{exp}}&=&0.5293(9)\times 10^{-2}/ps,\crn
	(\Delta  m_{B_d})_{\mathrm{exp}}&=&0.5065(19)/ps,\crn 
	(\Delta m_{B_s})_{\mathrm{exp}}&=&17.749(20)/ps. \label{meson-exp}
	\eea 
	However, due to the long-distance effect in $\Delta m_K$, the uncertainties in this system are considerable. Therefore, we require the theory to produce the data for the kaon mass difference within 30\%, namely    
	\bea 
	&& -0.3 <\fr{(\Delta m_{K})_{\text{NP}}}{(\Delta m_{K})_{\text{exp}}}< 0.3 .\label{constraint1}\eea
	The SM predictions for B-meson mass difference are more accurate than those of kaon, and we have the following constraints by combining quadrature of the relative errors in the SM predictions and measurements \cite{chang-he}
	
	\bea 
	0.6<\fr{(\Delta m_{B_d})_{\mathrm{exp}}}{(\Delta m_{B_d})_{\mathrm{SM}}}<1.17, \hs  0.71<\fr{(\Delta m_{B_s})_{\mathrm{exp}}}{(\Delta m_{B_s})_{\mathrm{SM}}}<1.2, \eea 
	or equivalently  
	\bea 
	-0.4<\fr{(\Delta m_{B_d})_{\text{NP}}}{(\Delta m_{B_d})_{\mathrm{SM}}}<0.17, \hs  -0.29<\fr{(\Delta m_{B_s})_{\text{NP}}}{(\Delta m_{B_s})_{\mathrm{SM}}}<0.2 . \label{constraint2} 
	\eea 	
	Let us do a numerical study from a set of all the input parameters that are taken by \cite{pdg,LQCD, Gambino:2016jkc,Charles:2015gya,Bona:2016dys}  
	\bea && m_d = 4.88(20) ,
	\hs m_s= 93.44(68), \hs m_b =4198(12), \hs m_t =172.4(7)\times 10^3,\crn
	&& f_K =155.7(3), \hs m_K=497.611(13),\hs f_{B_d}=190(1.3),\hs m_{B_d}=5279.65(12),\crn 
	&& f_{B_s}=230(1.3),\hs m_{B_s}=5366.88(14),\hs |(V_{\mathrm{CKM}})_{33}(V_{\mathrm{CKM}})_{31}^{*}|=0.0087(2), \crn &&  |(V_{\mathrm{CKM}})_{33}(V_{\mathrm{CKM}})_{32}^{*}/(V_{\mathrm{CKM}})_{23}|=0.982(1), \hs |(V_{\mathrm{CKM}})_{23}|=0.04200(64).  \eea 
	
	All mass parameters are in MeV. Besides, we assume $t_N=1,g=\sqrt{4\pi \al}/s_W$, where $\al=1/128$ and $s_W^2=0.231$. The mixing matrix for right-handed  quarks, $V_{u R}$, is a unitary matrix, whereas $V_{d R}$ is parameterized by three mixing angles, $\theta_{12}^R,\theta_{13}^R$ and $\theta_{23}^R$, as
	\bea
	\text{V}_{dR}&=& \left(%
	\begin{array}{ccc}
		c^R_{12}c^R_{23}-s^R_{12}s^R_{13}s^R_{23} &-s^R_{12}c^R_{13} &-c^R_{12}s^R_{23}-s^R_{12}s^R_{13}c^R_{23}\\
		
		s^R_{12}c^R_{23}+c^R_{12}s^R_{13}s^R_{23} &c^R_{12}c^R_{13} & -s^R_{12}s^R_{23}+c^R_{12}s^R_{13}c^R_{23}\\
		
		c^R_{13}s^R_{23} &-s^R_{13}&c^R_{13}c^R_{23}\\
	\end{array}%
	\right),
	\eea 
	where $s^R_{ig}=\sin \theta^R_{ij}$, $c^R_{if}=\cos \theta^R_{ij}$. For instance, we can choose $\theta_{12}^R=\pi/6,\theta_{13}^R=\pi/4$ and $\theta_{23}^R=\pi/3$. The NP scales require the following constraints $ w \sim \La \sim -f \gg u,v$, due to the condition of diagonalization for the mixing mass matrices in \cite{3311a}.

	\begin{figure}[H]% [H] is so declass\'e!
		\centering
		\begin{minipage}{0.5\textwidth}
			{\label{a}\includegraphics[width=\textwidth]{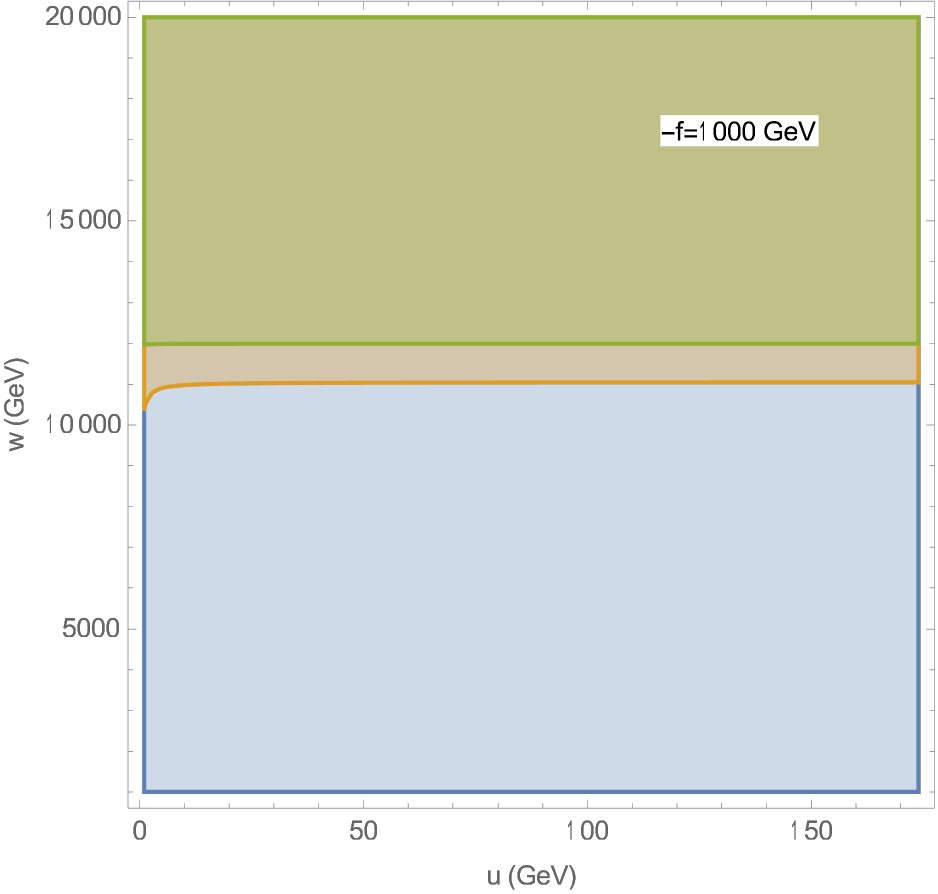}}
		\end{minipage}\hfill
		\begin{minipage}{0.5\textwidth}
			\includegraphics[width=\textwidth]{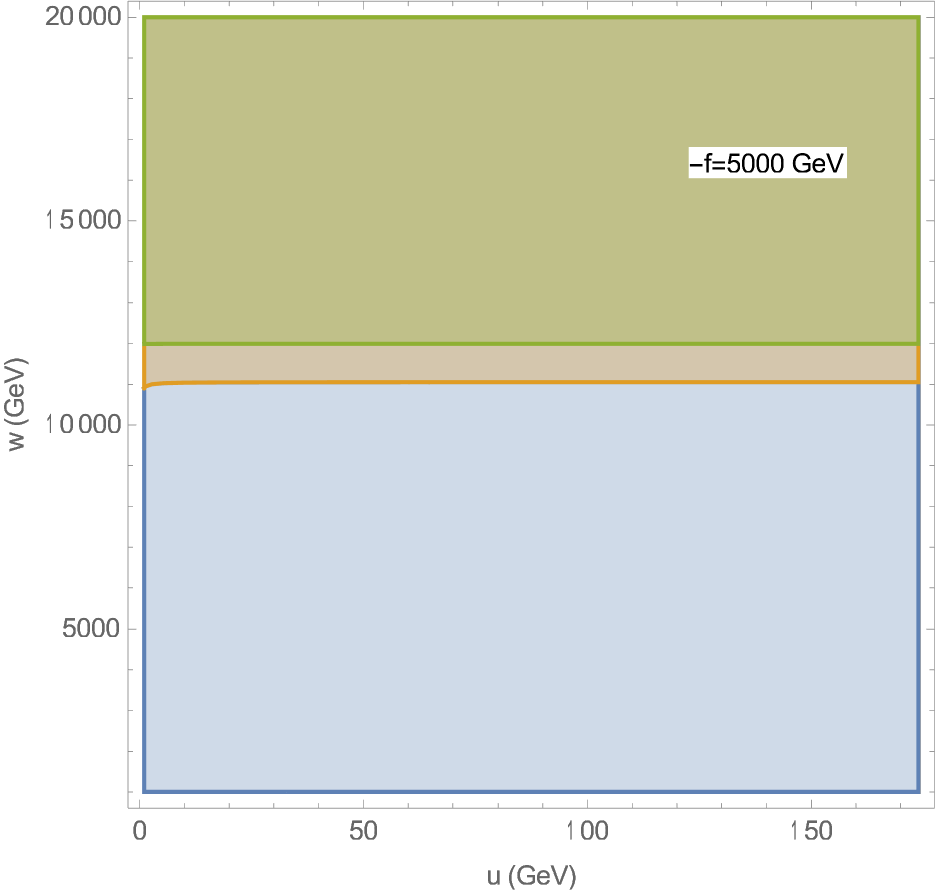}
		\end{minipage}\par
		\includegraphics[width=0.5\textwidth]{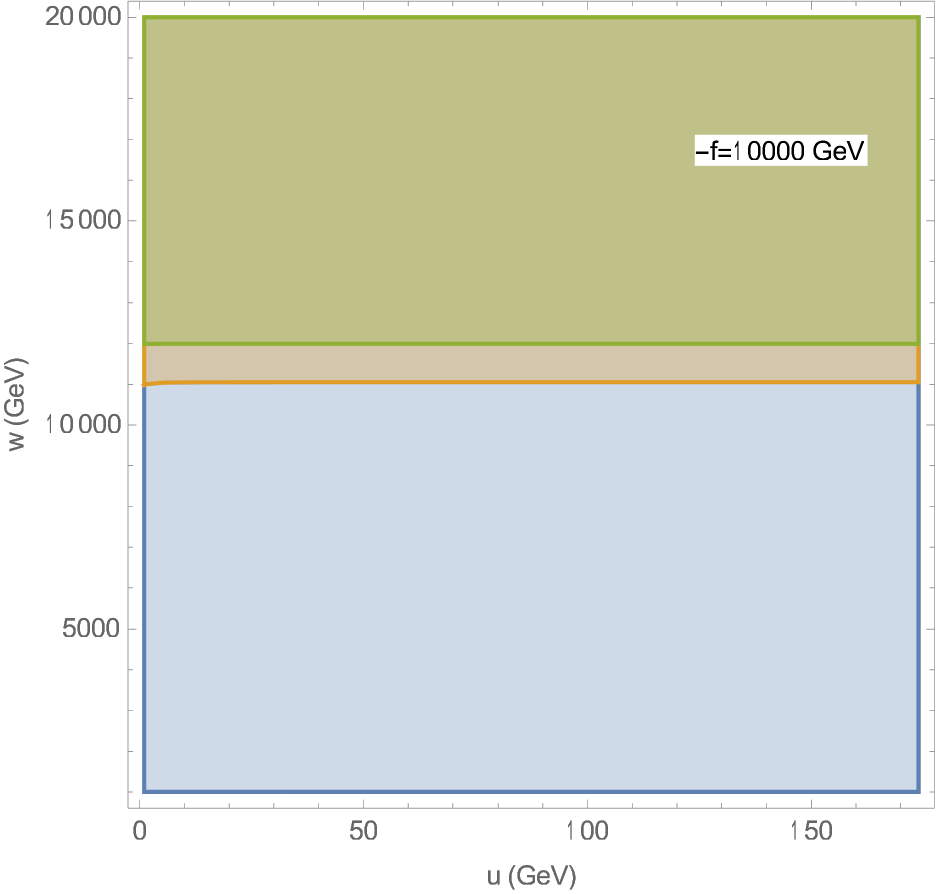}
		\caption{\label{Bsmixing} Constraints for $w$ and $u$ from the meson mass differences $\Delta m_{K}$,$\Delta m_{B_s}$ and $\Delta m_{B_d}$. The available region for $\Delta m_K$ is the whole frame, whereas the orange and green regions are for $\Delta m_{B_s}$ and $\Delta m_{B_d}.$}
	\end{figure}
	We first study the role of FCNCs coupled to the scalar fields, $H_1, \mathcal{A}$, in meson mixing parameters. To see its effect, we change the $f$-parameter, which only affects the masses of  the $H_1, \mathcal{A}$ (see in Eq. (\ref{m-scalar})). Specifically, in Fig. \ref{Bsmixing}, we draw contours of the mass differences $\Delta m_K$, $\Delta m_{B_s}$, and $\Delta m_{B_d}$, as functions of the NP scale $w$ and $u$ for three different choices of $f$-parameter as $f=-1000$ GeV, $f=-5000$ GeV and $f=-10000$ GeV. There are almost no differences between the three figures. That is, the mixing parameters are affected slightly by FCNCs coupled to the scalar fields.  
	
\begin{figure}[H]% [H] is so declass\'e!
		\centering
		\begin{minipage}{0.5\textwidth}
			{\label{a}\includegraphics[width=\textwidth]{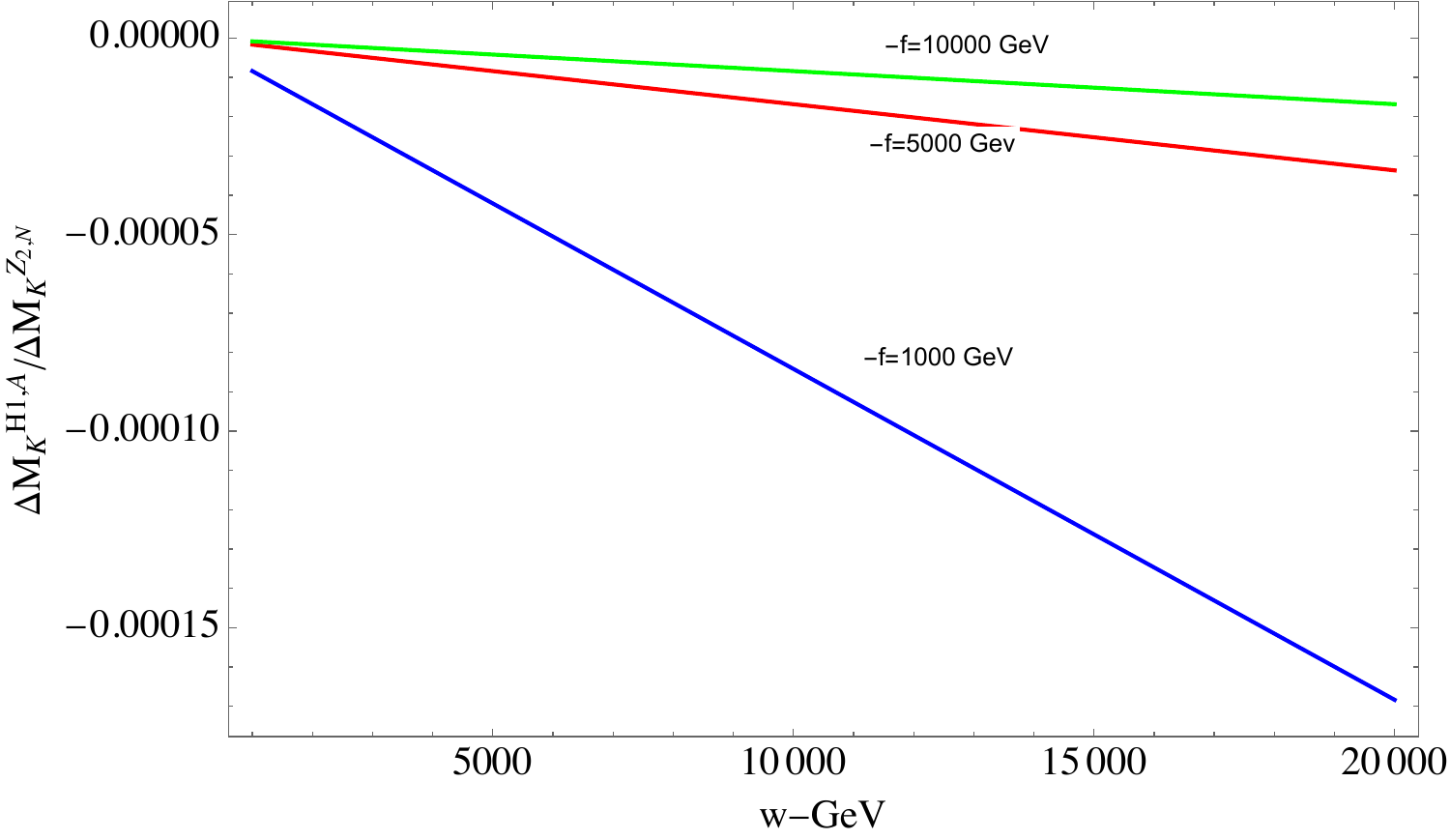}}
		\end{minipage}\hfill
		\begin{minipage}{0.5\textwidth}
			\includegraphics[width=\textwidth]{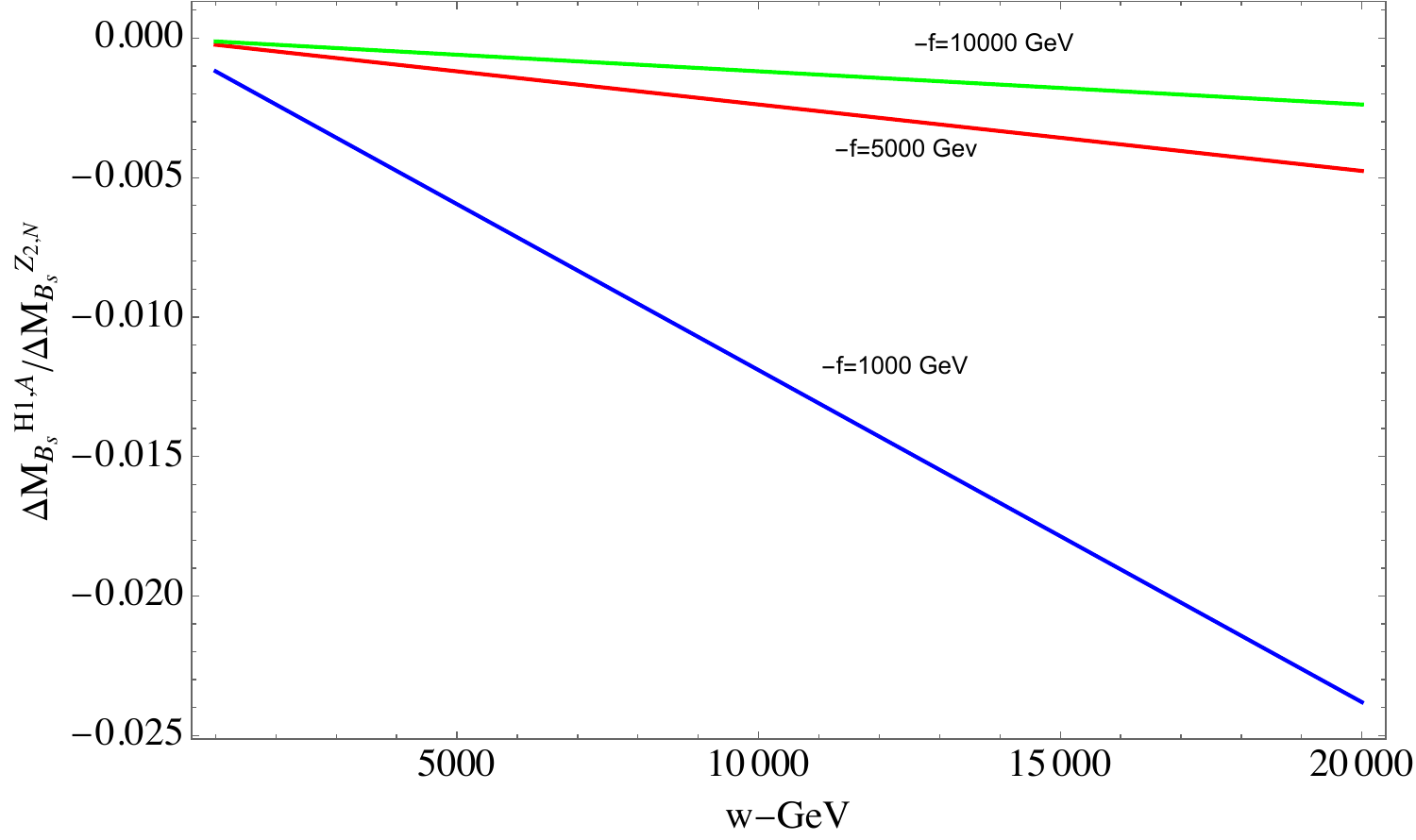}
		\end{minipage}\par
		\includegraphics[width=0.5\textwidth]{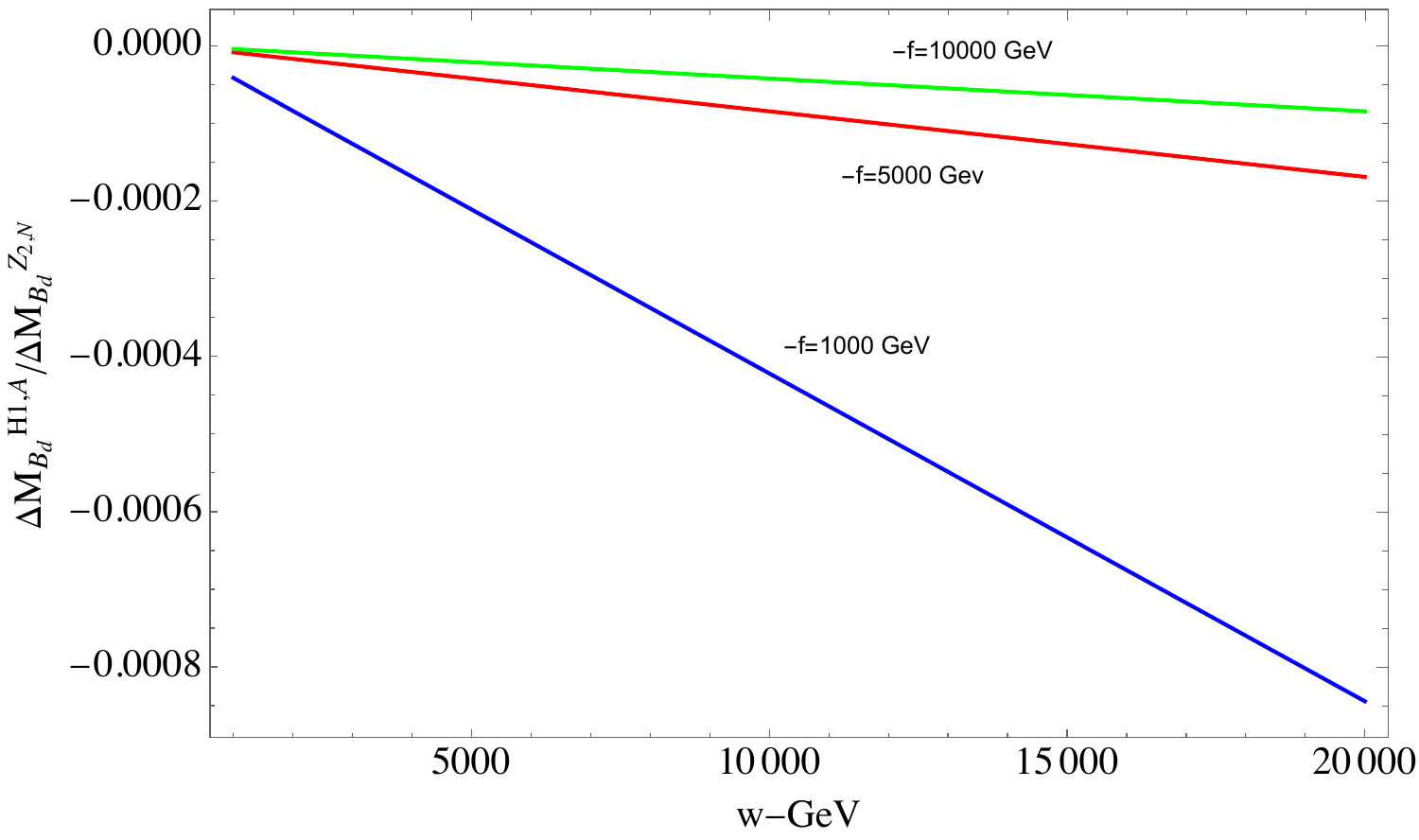}
		\caption{\label{HiggsVSGauge} The figures present the dependence of ratios $\Delta m_{K,B_s,B_d}^{H_1,A}/\Delta m_{K,B_s,B_d}^{Z_2,Z_N}$ on the NP scale $w$.}
	\end{figure}

	Next, we consider the contributions of FCNCs coupled to new gauge bosons to the meson mixing parameters. To estimate how important they are, we compare their contributions with those of the new scalar bosons. The ratio of these two contributions is presented in Fig. \ref{HiggsVSGauge}. The results show that the significant contribution comes from the FCNCs of new gauge bosons. It once again clarifies the small effect of the new scalar fields on the meson mixing systems. 
	
	Finally, we investigate the constraints on the VEVs from $\Delta m_{K, B_s, B_d}$. In Fig.\ref{Bsmixing}, the allowed region of parameters that satisfies the constraints given in Eqs. (\ref{constraint1}),(\ref{constraint2}) is the green one. The electroweak symmetry breaking energy scale, $u$, is not constrained by conditions imposed on the meson mass mixing parameters. However, these conditions affect the NP scale $w $. From Fig. \ref{Bsmixing}, we obtain a lower bound on the NP scale,  $w>$ 12 TeV. This lower bound is more stringent and is remarkably larger than that obtained previously \cite{3311a}. This difference is because, in the previous study, the authors compared the NP contributions with experimental values and ignored the SM contributions to the theoretical predictions.  Moreover, Eq. (131) in \cite{3311a}, the authors used $(\Delta m_{B_s})_{\text{NP}} < \frac{1}{(100 \ \text{TeV})^2} m_{B_s}f_{B_s}^2 \simeq 41.2871 /ps$, the upper limit for $(\Delta m_{B_s})_{\text{NP}}$ is even greater than that of the experimental value given in Eq. (\ref{meson-exp}). This is not reasonable because the theoretical prediction must consist of both SM and NP contributions. We must also consider the uncertainties of both SM and experimental predictions. Thus, the NP contributions have to be constrained by the conditions given in Eqs. (\ref{constraint1}, \ref{constraint2}).
	
	\subsection{$B_s \rightarrow \mu^+ \mu^-$, $B \rightarrow K^* \mu^+ \mu^-  \ \text{and} \  B^{+}\rightarrow K^{+}\mu^{+}\mu^{-} $}
	Rare decays of $B$ meson, in particular of the decay induced by the quark level transition, $B_s \rightarrow \mu^+ \mu^-$, $B\rightarrow K^* \mu^+ \mu^-$ and $B^{+}\rightarrow K^{+}\mu^{+}\mu^{-}$, are sensitive to physics beyond the SM. The NP effects can be quantified via the language of the effective theory. The effective Hamiltonian related to the above decays is determined by the quark FCNCs given in (\ref{quark1}), (\ref{quark2a}) and the lepton flavor-conserving neutral currents (LFCNCs). The LFCNCs coupled to the neutral scalars, $H_1, \mathcal{A}$, obtained from Eq. (\ref{yuklep1}) as follows
	\bea
	-\frac{g}{2 m_{W}}\frac{u}{v} \bar{l}^{\prime}_{aL} M^{lD}_{ab} l^\prime_{bR} (H_1+i\mathcal{A})+H.c.,
	\label{lepton1}\eea
	where $M^{lD} =\text{Diag}(m_e, m_\mu,m_\tau)$. It is worth noting that there is no neutral Higgs mediated FCNC in the lepton sector. The interactions of $Z_2$ and $Z_N$ with two charged leptons have been written in \cite{3311} read
	\bea
	-\frac{g}{2c_W} \bar{f}\ga^\mu \left(g_V^{Z_2}(f)-g_A^{Z_2}(f) \ga_5\right)f Z_{2 \mu}-\frac{g}{2c_W}\bar{f}\ga^\mu \left(g_V^{Z_N}(f) -g_{A}^{Z_N}(f) \ga_5 \right)f Z_{N \mu}
	\label{lepton2}, \eea
	where the form of coefficients $g_V^{Z_2,Z_N}, g_A^{Z_2,Z_N}$ are found in \cite{3311}.  
	
	Combining the quark FCNCs and the LFCNCs, we obtain the effective Hamiltonian for $B_s \rightarrow \mu^+ \mu^-$, $B \rightarrow K^* \mu^+ \mu^-$ and $B^{+}\rightarrow K^{+}\mu^{+}\mu^{-}$ processes as follows
	\bea
	\mathcal{H}_{\text{eff}}= -\fr{4G_F}{\sqrt{2}}V_{tb}V^*_{ts}\sum_{i=9,10,S,P}\left(C_i(\mu)\mathcal{O}_i(\mu)+C^\prime_i(\mu)\mathcal{O}^\prime_i(\mu)\right),
	\label{eff1}\eea
	where the operators are defined by
	\bea
	\mathcal{O}_9&& =\frac{e^2}{(4\pi)^2}(\bar{s}\ga_\mu P_L b)(\bar{l}\ga^\mu l), \hs \hs \hs \mathcal{O}_{10}=\frac{e^2}{(4\pi)^2}(\bar{s} \ga_\mu P_L b)(\bar{l} \ga^\mu \ga^5 l),  \\
	\mathcal{O}_S && =\frac{e^2}{(4 \pi)^2}(\bar{s}P_R b)(\bar{l}l), \hs \hs \hs  \hs \hs \mathcal{O}_P=\frac{e^2}{(4 \pi)^2}\left(\bar{s}P_R b \right) \left(\bar{l}\ga_5 l\right).
	\eea
	The operators $\mathcal{O}^\prime_{\text{9,10,S,P}}$ are obtained from $\mathcal{O}_{\text{9,10,S,P}}$ by replacing $P_L\leftrightarrow P_R$. Their Wilson coefficients consist of the SM leading and tree-level NP contributions. For $C_{9,10}$ we split into the SM and NP contributions as:  $C_{9,10}=C_{9,10}^{\text{SM}}+C_{9,10}^{\text{NP}}$, where the central points of $C_{9,10}^{\text{SM}}$ are given in \cite{SMC10}, $C_{10}^{\text{SM}}=-4.198, C_{9}^{\text{SM}}=4.344$, and the $C_{9,10,\text{S,P}}^{\text{NP}}$ are written by   
	\bea
	C_9^{\text{NP}}&& =-\Theta_{23}\fr{m_W^2}{c_W V_{tb}V_{ts}^*}\frac{(4\pi)^2}{e^2}\left(\fr{g_2}{g}\fr{g_V^{Z_2}(f)}{m_{Z_2}^2}+\fr{g_N}{g}\frac{g_V^{Z_N}(f)}{m_{Z_N}^2}\right), \crn
	C_{10}^{\text{NP}}&& =\Theta_{23}\fr{m_W^2}{c_W V_{tb}V_{ts}^*}\frac{(4\pi)^2}{e^2}\left(\fr{g_2}{g}\fr{g_A^{Z_2}(f)}{m_{Z_2}^2}+\fr{g_N}{g}\frac{g_A^{Z_N}(f)}{m_{Z_N}^2}\right). \eea
	Noting that $C_{\text{S,P}}^{\text{SM}}=C_{\text{S,P}}^{\prime \text{SM}}=0$. Therefore, the $C_{\text{S,P}},C^{'}_{\text{S,P}} $ are obtained by NP contributions as follows
	\bea
	C_S^{\text{NP}} && = \frac{8 \pi^2	}{e^2}\frac{1}{V_{tb}V_{ts}^*}\fr{\Ga^d_{23}\Ga^l_{\al \al}}{m_{H_1}^2}, \hs \hs \hs
	C_S^{\prime \text{NP}}  = \frac{8 \pi^2}{e^2}\frac{1}{V_{tb}V_{ts}^*}\fr{\left( \Ga^d_{32}\right)^*\Ga^l_{\al \al}}{m_{H_1}^2},\crn
	C_P^{\text{NP}}&&=-\frac{8 \pi^2}{e^2}\frac{1}{V_{tb}V_{ts}^*}\fr{\Ga^d_{23}\Delta^l_{\al \al}}{m_{A}^2}, \hs \hs \hs  C_P^{\prime \text{NP}}  = \frac{8 \pi^2}{e^2}\frac{1}{V_{tb}V_{ts}^*}\fr{\left( \Ga^d_{32}\right)^*\Delta^l_{\al \al}}{m_{A}^2},
	\eea
	where $\Ga^l_{\al \al}=\Delta^l_{\al a}=\fr{u}{v}m_{l_\al}$.

	From the effective Hamiltonian given in (\ref{eff1}), we obtain the branching ratio of the  $B_s \rightarrow l_{\al}^+ l_{\al}^-$ decay 
	\bea
	&& \text{Br}(B_s \rightarrow l_\al^+ l_\al^-)_{\text{theory}} = \frac{\tau_{B_s}}{64 \pi^3}\al^2 G_F^2 f_{B_s}^2|V_{tb}V^*_{ts}|^2 m_{B_s}
	\sqrt{1-\fr{4m_{l_\al}^2}{m_{B_s}^2}} \crn  && \times \left\{ \left(1-\frac{4m_{l_\al}^2}{m_{B_s}^2}\right)\left |\fr{m_{B_s}^2}{m_b+m_s} \left(C_S-C_S^\prime \right)\right |^2+\left|2m_{l_\al}\left(C_{10}-C_{10}^\prime \right) +\fr{m_{B_s}^2}{m_b+m_s}\left(C_P-C_P^\prime \right) \right|^2\right\} ,
	\eea
	where $\tau_{B_s}$ is the total lifetime of the $B_s$ meson. If including the effect of oscillations in the $B_s-\bar{B}_s$ system, the theoretical and experimental results are related by \cite{lienhe1} 
	\bea
	\text{Br}(B_s\rightarrow l_\al^+ l_\al^-)_{\text{exp}} \simeq \fr{1}{1-y_s}\text{Br}(B_s\rightarrow l_\al^+ l_\al^-)_{\text{theory}},
	\eea 
	where $y_s =\frac{\Delta \Gamma_{B_s}}{2\Gamma_{B_s}}= 0.0645(3)$ \cite{HFLAV}. For $B_s \rightarrow e^+e^-$, the SM prediction \cite{SMBsee} is
	\bea
	\text{Br}(B_s \rightarrow e^+ e^-)_{\text{SM}}= (8.54 \pm 0.55) \times 10^{-14},
	\eea
	and the experimental bound has been given in \cite{Ex-Bsee} as
	\bea
	\text{Br}(B_{s}\rightarrow e^+ e^-)_{\text{exp}} < 2.8 \times 10^{-7}.
	\eea 
	The SM contribution to the branching ratio of $B_s \rightarrow e^+ e^-$ is strongly suppressed to the current experimental upper bound. It may be an excellent place to look for NP. Completely contrary to $B_{s}\rightarrow e^+ e^-$, the very recent measurement of the branching ratio $(B_s\rightarrow \mu^{+}\mu^{-})$ is given by \cite{LHCb2021}
	\bea \text{Br}(B_s\rightarrow \mu^+ \mu^-)_{\text{exp}}=(3.09^{+0.46\ +0.15}_{-0.43 \ -0.11 } ) \times 10^{-9}.\eea   
	This experimental upper bound closes to the central value of the SM prediction (including the effect of $B_s-\bar{B}_s$ oscillations) that has been studied in \cite{SMBsmumu} 
		\bea \text{Br}\left(B_s \rightarrow \mu^+ \mu^- \right)_{\text{SM}}=\left(3.66 \pm 0.14 \right) \times 10^{-9}.\eea  It shows that experimental results are in slight tension with the SM prediction of Br$(B_s \rightarrow \mu^+ \mu^-)$. NP effects in $B_s \rightarrow \mu^+ \mu^- $ lead to new stringent constraints on NP scale. Let us concentrate on the numerical study of $B_s \rightarrow \mu^+ \mu^-$.  
	
	\begin{figure}[H]
		\centering
		\begin{tabular}{cc}
			\includegraphics[width=8cm]{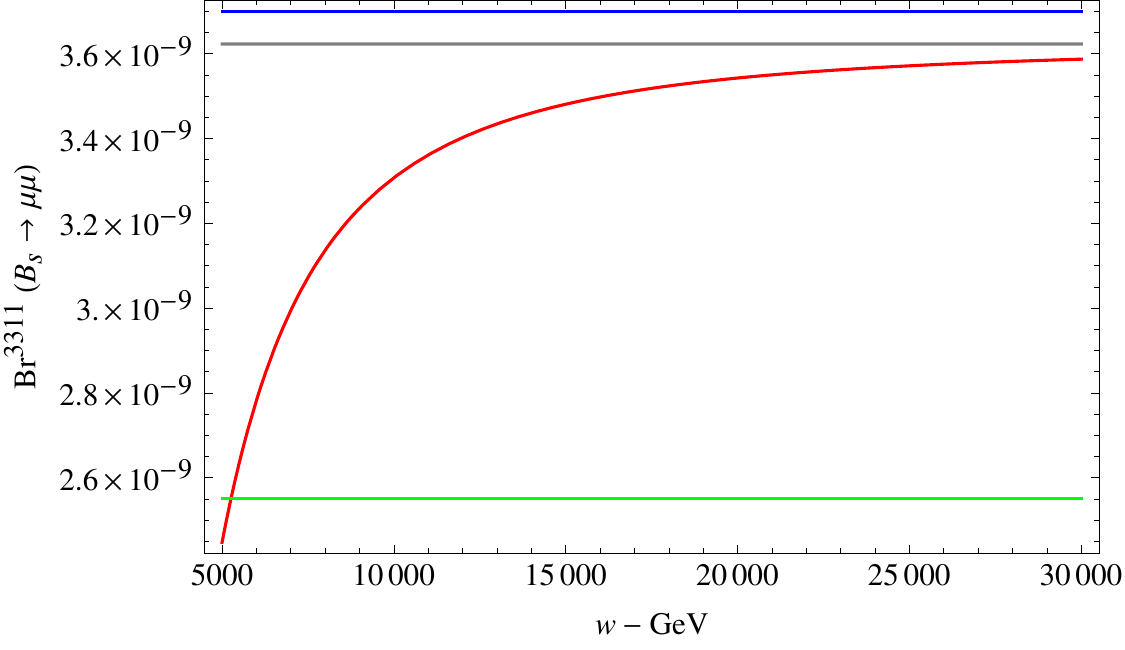}&
			\includegraphics[width=8cm]{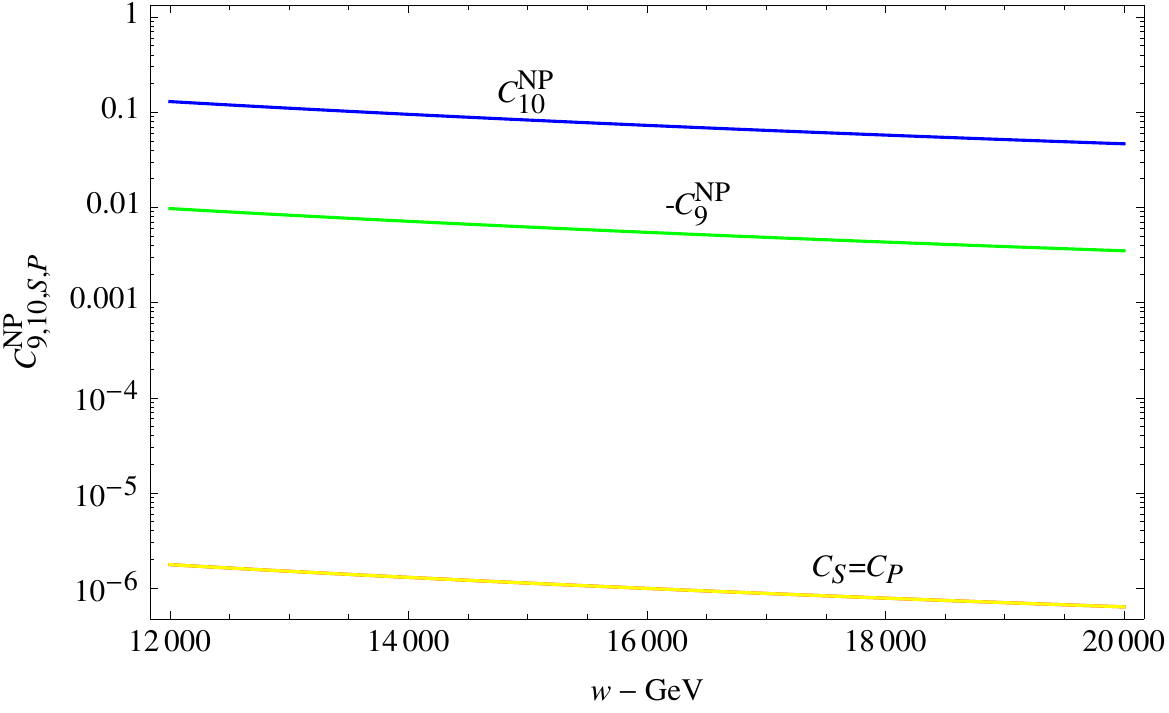}
		\end{tabular}
		\caption{\label{branching} The left panel draws the  $\text{Br}(B_s \rightarrow \mu^{+} \mu^{-})$: red curve  presents  the prediction values of the 3-3-1-1 model, gray line represents the central values of the SM prediction. The blue and green lines represent the experimental upper and lower bounds.  The right panel predicts the NP contributions to the Wilson coefficients. Here both panels are plotted by fixing: $\La=1000w,f=-w,u=200 \ \text{GeV}$. Other parameters are selected as done in the Sec. \ref{FCNC-treelevel}}
	\end{figure}
	In Fig. \ref{branching}, the red curve in the left panel demonstrates the $\text{Br}(B_s \rightarrow \mu^{+} \mu^{-})$ in the $\text{3-3-1-1}$ model as a function of the new symmetry breaking scale. The predicted results are only consistent with the current experimental bounds if the VEV, $w$, is larger than $\text{5 TeV}$. This bound is not as strict as the constraints obtained from studying the meson oscillations in Sec. \ref{BBmix}. So, the best fit region pulls for both $(\bar{B}_s-B_s)$ mixing and $\text{Br}(B_s \rightarrow \mu^+ \mu^-)$ experimental bounds is $w> 12$ TeV. In the right panel of Fig. \ref{branching}, we draw the NP contributions to each Wilson coefficient. Compared to the $C_{9,10}^{\text{NP}}$, the $C_{\text{S,P}}$ are further suppressed by a factor of $10^{-4} \div 10^{-5}$. So, the main contribution of the NP to the $\text{Br}(B_s \rightarrow \mu^{+} \mu^{-})$ comes from the $C_{10}^{\text{NP}}$. In the limit $w > \text{12 TeV} $, the $C_{10}^{\text{NP}}$ is positive. It causes the $\text{Br}(B_s \rightarrow \mu^{+} \mu^{-})$ reduced about $5 \%$ ,  which brings the theoretical prediction and experimental values get closer together.

	%We would like to note that the corrections from new physics to the Wilson coefficients, $ C_{9}, C_{10}$ also effects %in the $b \rightarrow s \mu^+ \mu^-$ transitions. 
	If the $C_{10}^{\text{NP}}$ affects  the decay process $B_s \rightarrow \mu^+ \mu^-$, the $C_{9}^{\text{NP}}$ plays a crucial role in $B \rightarrow K^* \mu^+ \mu^-$ decay. The current experimental measurements of the $b \rightarrow s \mu^+ \mu^-$ have attracted and led to many model-independent global analyses \cite{bsll,bsll1,Alok:2019ufo,Alguero:2021anc,Geng:2021nhg,Altmannshofer:2021qrr,Hurth:2020ehu,Cornella:2021sby} assuming the presence of \text{NP}. The anomalies of the $B \rightarrow K^* \mu^+ \mu^-$ decay were explained if there exists a large negative contribution to the Wilson coefficient $C_{9}^{\text{NP}}$. The best-fit point for the $C_9^{\text{NP}}$ varies around $-1.1$. The green line in the right panel of Fig. \ref{branching} predicts the $C_9^{\text{NP}}$ in the $\text{3-3-1-1}$ model. In the limit, $w> 12 \ \text{TeV}$, we obtain its maximal prediction value $C_9^{\text{NP}} \simeq -0.01$. So, the \text{NP} coming from the $\text{3-3-1-1}$ model can not explain the anomalies of $B \rightarrow K^* \mu^+ \mu^-$ process. 
	
	The measurements of the branching fraction of the decay $B^+\rightarrow K^+ \mu^+ \mu^-$  \cite{bsll2,bsll2-bs2} have turned out to be slightly on the low side compared to SM expectations.  Both the $C_9, C_{10}$ contribute to the $\text{Br}\left(B^+\rightarrow K^+ \mu^+ \mu^-\right)$. As predicted by the 3-3-1-1 model, the NP contribution to these parameters is minimal (see Fig. \ref{branching}) because the NP scale satisfies the constraint $w >12 \ \text{TeV}$.  Both the $C_{9}^{\text{NP}}$ and $C_{10}^{\text{NP}}$ are too low and far from  the values of global analysis, see in  \cite{bsll, bsll1, Alok:2019ufo,Alguero:2021anc}.  Thus, we believe that the NP effects in $B^+ \rightarrow K^+ \mu^+ \mu^-$ remain small in  the 3-3-1-1 model.
	\section{\label{Radiative}Radiative processes}
	\subsection{\label{b-sgamma}$ b \rightarrow s \gamma$ decay}
	The branching fraction and the photon energy spectrum of the radiative penguin $b \rightarrow \text{s} \gamma$ process have been firstly reported by \text{CLEO} experiment, \text{Br}$(b \rightarrow s \gamma)=(3.21\pm 0.43 \pm 0.27^{+0.18}_{-0.10}) \times 10^{-4}$ \cite{bsgamma}. Recently, \text{HFLAV} group has obtained the average result by combining the measurements from CLEO, BaBar and Belle, \text{Br}$(b \rightarrow s \gamma)=(3.32\pm 0.15) \times 10^{-4}$ \cite{HFLAV} for a photon-energy cut-off $E_{\gamma}>1.6$ GeV. This result is in good agreement with the SM prediction up to Next-to-Next-to-Leading Order (NNLO) Br$(b \rightarrow s \gamma)=(3.36 \pm 0.23) \times 10^{-4}$ \cite{bsg-SM},\cite{bsg-SM-1}, with the same energy cut-off $E_{\gamma}$. It suggests that the \text{NP} contributions to this process, if any, have to be small. Thus, studying the $b\rightarrow s \gamma$ decay can give a strong constraint on the \text{NP} scale. The radiative process $b \rightarrow s \gamma$ is most conveniently described in the framework of an effective theory that arises after decoupling of new particles. Excluding the charged currents associated with the $W_{\mu}^\pm$ gauge boson, the $\text{3-3-1-1}$ model contains new charged currents, which couple to the new charged gauge bosons $Y_\mu^\pm$, two charged Higgs bosons $H_4^\pm, H_5^\pm$, and the $\text{FCNCs}$ coupled to the $Z_{2,N}$ as given in Eq. (\ref{quark2a}). All of the above currents generate the $b \rightarrow s \gamma$ process.
	
	%due to the \text{FCNCs} given in Eq. (\ref{quark2a}).
	%We will derive the limits on new physics scales of our model by studying the $\text{b}\rightarrow \text{s} \gamma$ amplitude. 
	Let us write down the charged scalar currents related to $b \rightarrow s \gamma$. The $H_4^\pm$ only couples to the exotic quarks, so it does not create the flavor-changing charged currents (\text{FCCCs}) for SM quarks. While $H_5^{\pm}$ couples to the SM quarks and creates the scalar \text{FCCCs}. The relevant Lagrangian is
	\bea
	\mathcal{L}_{\text{Yukawa}}^{H_5^\pm}=\fr{g}{\sqrt{2}m_W}\left\{\bar{d}^{\prime}_L\mathcal{X} \mathcal{M}_uu^\prime_R+\bar{d}^{\prime}_R\mathcal{M}_d \mathcal{Y}  u^\prime_L\right\} H_5^{-} +H.c.,
	\label{bs1}\eea
	where $\mathcal{Y}=t_\beta V_{\text{CKM}}^{\dagger} -\fr{2}{s_{2\beta}}\mathcal{T}$ and $\mathcal{X}=\fr{1}{t_\beta} V_{\text{CKM}}^{\dagger} -\fr{2}{s_ {2\beta}}\mathcal{T}$. The $\mathcal{T}$ is defined as $ \mathcal{T}_{ij}=(V_{d_L}^\dag)_{i3}(V_{u_L})_{3j}$, $s_{2\beta}=\sin 2\beta, t_{2\beta}=\tan 2\beta$.
	The charged currents associated with the $W^\pm, Y^\pm$, are described by the $\text{V-A}$ currents as follows
	\bea
	\mathcal{L}_{W,Y}^{\text{quark}}&&= \fr{g}{2\sqrt{2}}\bar{u}^{\prime}\ga^\mu (1-\ga_5)W^+_\mu V_{\text{CKM}}d^\prime+H.c. \crn &&+
	\fr{g}{2\sqrt{2}}\left\{\bar{d}^{\prime}_j (V_{d_L}^*)_{j3}\ga^\mu(1-\ga_5)Y^-_\mu U +\bar{D}_{\al}\ga^\mu (1-\ga_5)Y^-_\mu (V_{u_L})_{\al j}u^\prime_{j}\right\}+H.c. .
	\label{bs2}\eea
	The effective Hamiltonian for the decay $b \rightarrow s \gamma$  is 
	\bea \mathcal{H}_{\text{eff}}^{b \rightarrow s \gamma}&&=-\fr{4G_F}{\sqrt{2}}V_{tb}V_{ts}^{*}[C_7(\mu_b)\mathcal{O}_7+C_8(\mu_b)\mathcal{O}_8 + C_7'(\mu_b)\mathcal{O}'_7+C_8'(\mu_b)\mathcal{O}'_8],\eea  
	with $\mu_b=\mathcal{O}(m_b)$. The electromagnetic and chromomagnetic dipole operators $\mathcal{O}_7,\mathcal{O}_8$ are defined as 
	\bea
	\mathcal{O}_7=\fr{e}{(4\pi)^2} m_b(\bar{s}_{\al}\sigma_{\mu \nu}P_R b_{\al})F^{\mu \nu}, \hs \mathcal{O}_8=\fr{g_s}{(4\pi)^2} m_b(\bar{s}_{\al}\sigma_{\mu \nu}T^a_{\al \beta}P_R b_{\beta})G^{a \mu \nu},
	\eea
	and the primed operators $\mathcal{O}_{7,8}'$ are obtained by replacing $P_L \leftrightarrow P_R$. The Wilson coefficients $C_{7,8}(\mu_b)$ split as the sum of the SM and \text{3-3-1-1} contributions
	\bea
	C_{7,8}(\mu_b)=C_{7,8}^{\text{SM}}(\mu_b)+C_{7,8}^{\text{NP}}(\mu_b).  \label{bsgamma}
	\eea   
	Note that the Wilson coefficients $C_{7,8}'$ will be ignored in our calculation since they are suppressed by the ratio $m_s/m_b$. The SM Wilson coefficients $C_{7,8}^{\text{SM}}$ at the scale $\mu \sim m_W$ are first given by \cite{inami-lim} 
	\bea
	&& C^{\text{SM}(0)}_7(m_W)=\fr{m_t^2}{m_W^2}f_{\gamma}\left(\fr{m_t^2}{m_W^2}\right), \hs C_{8}^{\text{SM}(0)}(m_W )= \fr{m_t^2}{m_W^2}f_g \left(\fr{m_t^2}{m_W^2}\right), \eea
	where the index \text{0} indicates that the Wilson coefficients are calculated without QCD correction.
	
	The \text{NP} contributes to $C_{7,8}^{\text{NP}}$ at the quantum level via the higher order charged current interactions in Eqs. (\ref{bs1}), (\ref{bs2}) and the \text{FCNCs} given in Eq. (\ref{quark2a}). They can be split into each contribution as follows
	%the results  
	%contains contributions from the new charged Higgs boson $H_5^{\pm}$, new charged gauge boson $Y^{\pm}$ and new %neutral gauge bosons $Z_{2,N}$
	\bea C_{7,8}^{\text{NP}(0)}=C_{7,8}^{H_5(0)}(m_{H_5})+C_{7,8}^{Y(0)}(m_Y)+C_{7,8}^{Z_{2,N}(0)}(m_{Z_{2,N}}), \eea
	where
	\bea  
	&&  C_{7}^{H_5(0)}(m_{H_5})=\fr{m_t^2}{m_{H_5}^2}\left[\fr{1}{3} t_{\beta}^2 f_{\gamma}\left(\fr{m_t^2}{m_{H_5}^2}\right)+f^{\prime}_{\gamma} \left(\fr{m_t^2}{m_{H_5}^2}\right)\right], \crn
	&& C_{8}^{H_5(0)}(m_{H_5} )= \fr{m_t^2}{m_{H_5}^2}\left[ \fr{1}{3}t_{\beta}^2 f_{g}\left(\fr{m_t^2}{m_{H_5}^2}\right)+f^{\prime}_{g} \left(\fr{m_t^2}{m_{H_5}^2}\right)\right], \crn 
	&&  C_{7}^{Y(0)}(m_Y)=\fr{m_W^2}{m_Y^2}\fr{m_U^2}{m_Y^2}f_{\gamma}\left(\fr{m_U^2}{m_Y^2}\right), \hs C_{8}^{Y(0)}(m_Y )=\fr{m_W^2}{m_Y^2} \fr{m_U^2}{m_Y^2}f_g \left(\fr{m_U^2}{m_Y^2}\right), \eea
	with all functions $f_{\ga,g}$ and $f'_{\ga,g}$ are defined as shown below
	\bea
	&& f_{\gamma}(x)=\fr{(7-5x-8x^2)}{24(x-1)^3}+\fr{x(3x-2)}{4(x-1)^4} \ln{x}, \hs 
	f^{\prime}_{\gamma}(x)=\fr{(3-5x)}{12(x-1)^2}+\fr{(3x-2)}{6(x-1)^3}\ln{x}, \crn 
	&& f_{g}(x)=\fr{2+5x-x^2}{8(x-1)^3}-\fr{3x}{4(x-1)^4}\ln{x}, \hs 
	f^{\prime}_{g}(x)=\fr{3-x}{4(x-1)^2}-\fr{1}{2(x-1)^3}\ln{x}. 
	\eea
	The $C_{7}^{Z_{2,N}(0)}(m_{Z_{2,N}})$ are obtained by the \text{FCNCs} coupled to the $Z_{2,N}$ and have a form as given in \cite{buras-331bsgamma}
	\bea
	&&  C_{7}^{Z_{2,N}(0)}(m_{Z_{2,N}})=-\fr{2}{9g^2}\fr{m_W^2}{m_{Z_{2,N}}^2}\sum_{f=d,s,b}\fr{g^{fs*}_{L}g^{fb}_{L}}{V_{ts}^{*}V_{tb}}+\fr{2}{3g^2}\fr{m_W^2}{m_{Z_{2,N}}^2}\sum_{f=d,s,b}\fr{m_f}{m_b}\fr{g^{fs*}_{L}g^{fb}_{R}}{V_{ts}^{*}V_{tb}}, \crn 
	&& C_{8}^{Z_{2,N}(0)}(m_{Z_{2,N}} )= -3C_7^{Z_{2,N}}(m_{Z_{2,N}}) \eea 
	with $g^{ff}_{L,R}=[g_V^{Z_{2,N}}(f)\pm g_A^{Z_{2,N}}(f)]/2$ are the flavor-conversing couplings given in \cite{3311} while $g^{fs,fb}$ are the flavor-violating couplings defined in  Eq. (\ref{quark2a}).

	Noting that \text{QCD} corrections to $b \rightarrow s \gamma$ are important and have to be included to complete the analysis. The Ref. \cite{misiak-steinhauser} predicted $C_{7,8}^{\text{SM}}$ up to NNLO, $C_7^{\text{SM}}(\mu_b)=-0.3523$ for $\mu_b=2.5$ GeV. The recent calculations of the \text{NP} contributions to the $C_{7, 8}^{\text{NP}}$ have been considered at the Leading Order (\text{LO}) \cite{buras-331bsgamma}, \cite{buras-NPnumber}. In the following work, we study the effect of \text{QCD} corrections on the $C_{7,8}^{\text{NP}}$ at the \text{LO}. In the $\text{3-3-1-1}$ model, there are four heavy scales:  $m_Y$, $m_{Z_{2,N}}$ and $m_{H_5}$ . The difference between these scales can be ignored because the effects of QCD running are less important at high energies. Hence, we assume all calculations are at the same scale. For instance, we choose  $\mu\sim m_Y$. The \text{QCD} corrections for $C^{Z_{2,N}}_{7}$ are given by
	\bea
	C_{7}^{Z_{2,N}}(\mu_b)=\ka_7C_7^{Z_{2,N}}(m_Y)+\ka_8 C_8^{Z_{2,N}}(m_Y)+\Delta_{Z_{Z_{2,N}}}(\mu_b), \eea
	where $\ka_{7,8}$ are  \text{NP} magic numbers $\ka_7=0.39,\ka_8=0.130 $ at $\mu\sim 10$ TeV \cite{buras-NPnumber}. $\Delta_{Z_{2,N}}(\mu_b)$ are the contributions coming from the mixing of new neutral current-current operators, generated by the exchange of $Z_{2,N}$ with the dipole operators $\mathcal{O}_{7,8}$
	\bea && \Delta_{Z_{Z_{2,N}}}(\mu_b)=\sum_{A=L,R, \atop  f=u,c,t,d,s,b}\ka^f_{LA}\Delta_{LA}C_2^f(w)+\sum_{A=L,R}\hat{\ka}^d_{LA}\Delta_{LA}\hat{C}^d_2(w),\crn
	&& \Delta_{LA}C_2^f(m_Y)=-\fr{2}{g^2}\fr{g_L^{sb*}g_A^{ff}}{V_{ts}^{*}V_{tb}}, \hs  \Delta_{LA}\hat{C}_2^d(m_Y)=-\fr{2}{g^2}\fr{g_L^{sd*}g_A^{bd}}{V_{ts}^{*}V_{tb}}
	\eea 
	For $w=10$ TeV, we have $m_Y\simeq 3.2$ TeV, and obtain $C_{7}^{Z_{2,N}}(\mu_b)\simeq \mathcal{O}(10^{-5})$, which is strongly suppressed by the SM prediction, $C_7^{\text{SM}}(\mu_b)=-0.3523$. Therefore, in the next calculation, $C_{7}^{Z_{2,N}}$ can be ignored. If including the \text{LO} of \text{QCD} corrections, the $C_{7}^{Y}$ and $C_7^{H_5}$  have the form as \cite{buras-331bsgamma}, \cite{buras-NPnumber} 
	\bea 
	&& C_{7}^{Y}(\mu_b)=\ka_7C_7^{Y}(m_Y)+\ka_8 C_8^{Y}(m_Y), \crn
	&& C_{7}^{H_5}(\mu_b)=\ka_7C_7^{H_5}(m_Y)+\ka_8 C_8^{H_5}(m_Y). \eea 
	The branching ratio Br$(b \rightarrow s \gamma)$ is given as 
	\bea
	\text{Br}(b \rightarrow s \gamma) &=&\fr{6\al}{\pi C}\fr{|V_{ts}^{*}V_{tb}|^2}{|V_{cb}|^2}(|C_7(\mu_b)|^2+N(E_{\gamma}))\text{Br}(b\rightarrow c e\bar{\nu}_e) , \label{bra1}
	\eea 
	where $N(E_{\gamma})=3.6(6)\times 10^{-3}$ is a non-perturbative contribution,  $C=|V_{ub}/V_{cb}|^2\Ga(b\rightarrow ce\bar{\nu}_e)/\Ga(b\rightarrow ue\bar{\nu}_e)=0.580(16)$  \cite{misiak-steinhauser} and branching ratio for semi-leptonic decay  Br$(b\rightarrow ce\bar{\nu}_e)=0.1086(35)$ \cite{pdg}. Other parameters are input as in Sec. \ref{BBmix}.
	
	The \text{Br}$(b \rightarrow s \gamma)$ behaves as a function of the new particle masses, such as $m_{Y}, m_{H_5}, m_{U}$. These masses are understood as free parameters. In the limit, $u,v \ll -f\fr{u^2+v^2}{uv} \sim w \sim \La$, they can be rewritten as  
	\bea
	m_Y^2 \simeq \fr{g^2w^2}{4}, \hs  m_{H_5}^2 \simeq \fr{w^2}{\sqrt{2}}, \hs m_{U} = -\fr{h^Uw}{\sqrt{2}},
	\eea
	where, $g=\sqrt{4\pi \al / s_W^2} \simeq 0.63$, $h^U$ is unknown parameter. So, $m_U$ is arbitrary at the \text{TeV} energy scale, which can be higher or smaller than two other masses, $m_{H_5}, m_Y$. Without loss of generality, we investigate the mass hierarchy of new particles according to three scenarios: $ m_{H_5} > m_{Y} > m_U$,  $ m_{H_5} > m_{U} > m_Y$,  and $ m_{U} > m_{H_5} >m_Y$.

	\begin{figure}[H]% [H] is so declass\'e!
		\centering
		\begin{minipage}{0.5\textwidth}
			{\label{a}\includegraphics[width=\textwidth]{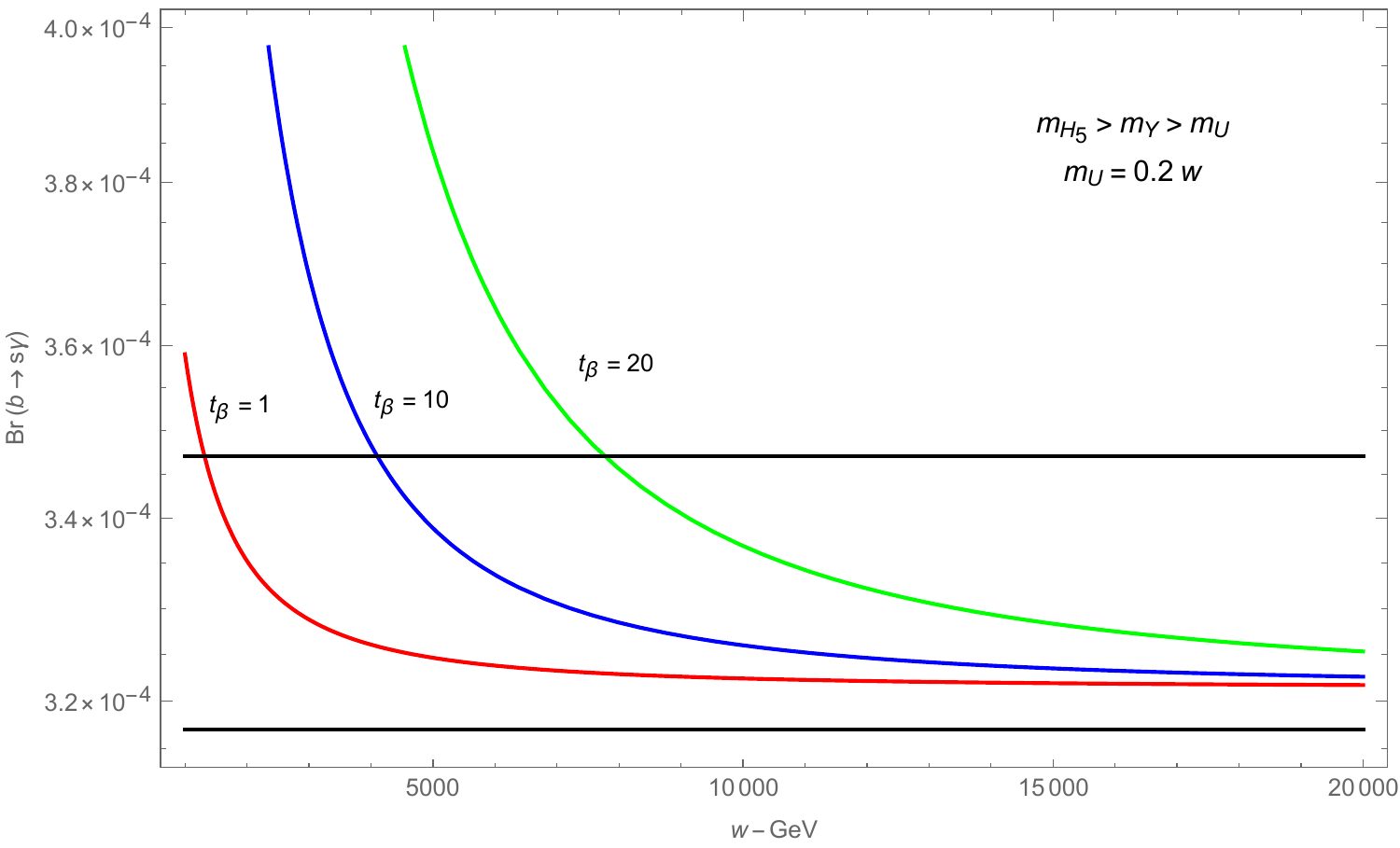}}
		\end{minipage}\hfill
		\begin{minipage}{0.5\textwidth}
			\includegraphics[width=\textwidth]{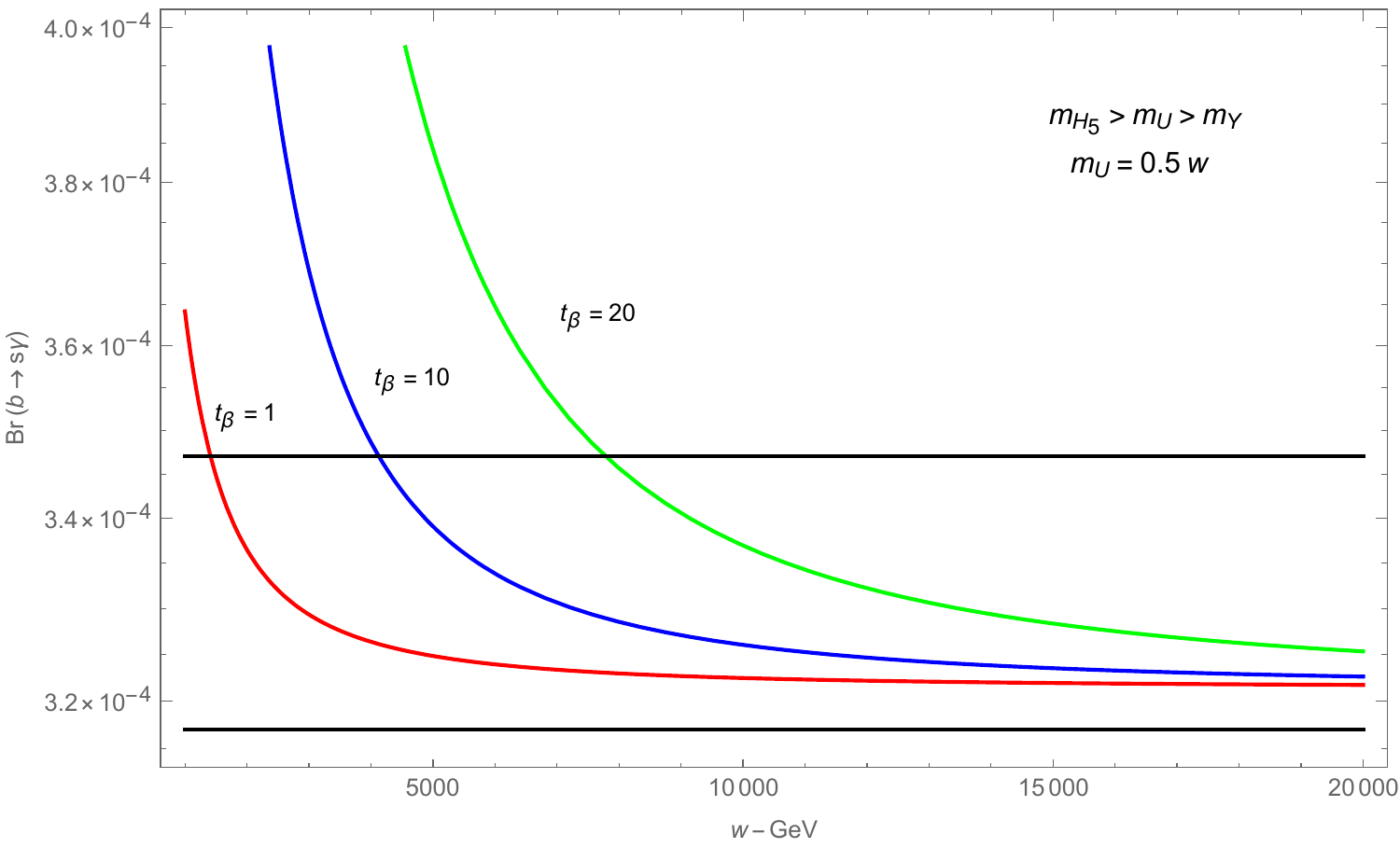}
		\end{minipage}\par
		\includegraphics[width=0.5\textwidth]{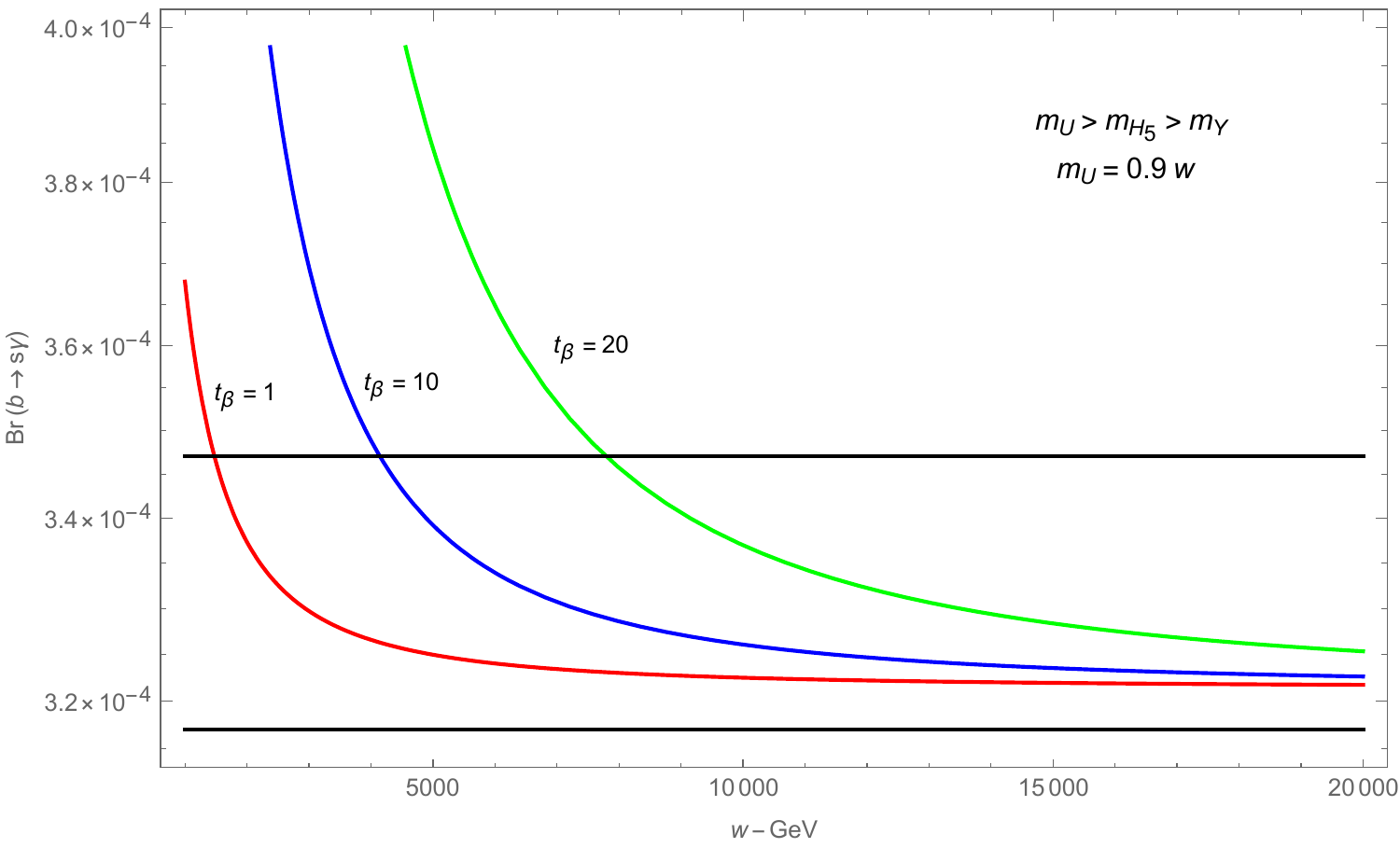}
		\caption{\label{plot-mY,H5} The dependence of the \text{Br}$(b \rightarrow s \gamma)$ on the \text{NP} scale $w$ in the limit, $u,v \ll -f\fr{u^2+v^2}{uv} \sim w \sim \La$. The solid black lines indicate the current experimental constraint Br$(b\rightarrow s\gamma)=(3.32\pm0.15)\times 10^{-4}$ \cite{HFLAV}}.
	\end{figure}
	
	In Fig. \ref{plot-mY,H5}, we show the dependence of \text{Br}$(b \rightarrow s \gamma)$ on the NP scale $w$ in the limit $u,v \ll -f\fr{u^2+v^2}{uv} \sim w \sim \La$. Each panel corresponds to the scenarios of mass hierarchy and three different choices of $t_\beta$. We see that the branching ratio strongly depends on the values of $t_\beta$ where the term containing $t_\beta$ comes from $C_7^{H_5}$. So we conclude that $C_7^{H_5}$ plays an important role in the radiative decay process $b \rightarrow s \gamma$. This is true for all three scenarios of the mass hierarchy. Besides, Fig. \ref{plot-mY,H5} indicates that the mass hierarchy does not affect Br$(b \rightarrow s \gamma)$ much. This result is understood as the main contribution coming from $C_7^{H_5}$, and it is stronger than other contributions by the coefficient $t_\beta^2$. In the large $t_\beta$ limit, the \text{Br}$(b \rightarrow s \gamma) \simeq |C_7^{H_5}|^2 \simeq \frac{t_\beta^2}{w^2}$. The lower bound on the \text{NP} scale depends on the value of the $t_\beta$, specifically, $w \ge 1 \ \text{TeV}$ for $t_\beta=1$; $w \ge 4.1 \ \text{TeV}$ for $t_\beta =10$; $w \ge 7.7 \ \text{TeV}$ for $t_\beta =20$. These limits are weaker than the ones mentioned above. 
	
	To close this section, we consider the influence of \text{NP} on the $\text{Br}(b\rightarrow s \gamma)$ in the limit  $u,v \ll -f \sim w \sim \La$. In Fig. \ref{plot-mY,H5-case II}, we see that the dependence of branching ratio on $t_\beta$ is not as strong as predicted in Fig. \ref{plot-mY,H5}. This difference can be explained by the dependence of $m_{H_5}$ on $t_{\beta}$, $m_{H_5}= 0.85 w\left(t_{\beta}+\frac{1}{t_\beta}\right)$. Therefore, \text{Br}$(b \rightarrow s \gamma) \simeq |C_7^{H_5}|^2 \simeq t_{\beta}^2\frac{1}{m_{H_5}^2}\simeq t_{\beta}\fr{1}{w^2}$, whereas  Br$(b \rightarrow s \gamma) \simeq t_{\beta}^2\fr{1}{w^2}$ for the previous case. This leads to the lower limit of the \text{NP} also changing for each choice of $t_\beta$. In the limit given in Sec. \ref{BBmix}, $w > 12 \ \text{TeV}$,  the affect of $t_{\beta}$ to \text{Br}$(b\rightarrow s\gamma)$ becomes trivial and the predicted branching ratio approaches the central value of the experimental bounds. 
	\begin{figure}[H]% [H] is so declass\'e!
		\centering
		\begin{minipage}{0.5\textwidth}
			{\label{a}\includegraphics[width=\textwidth]{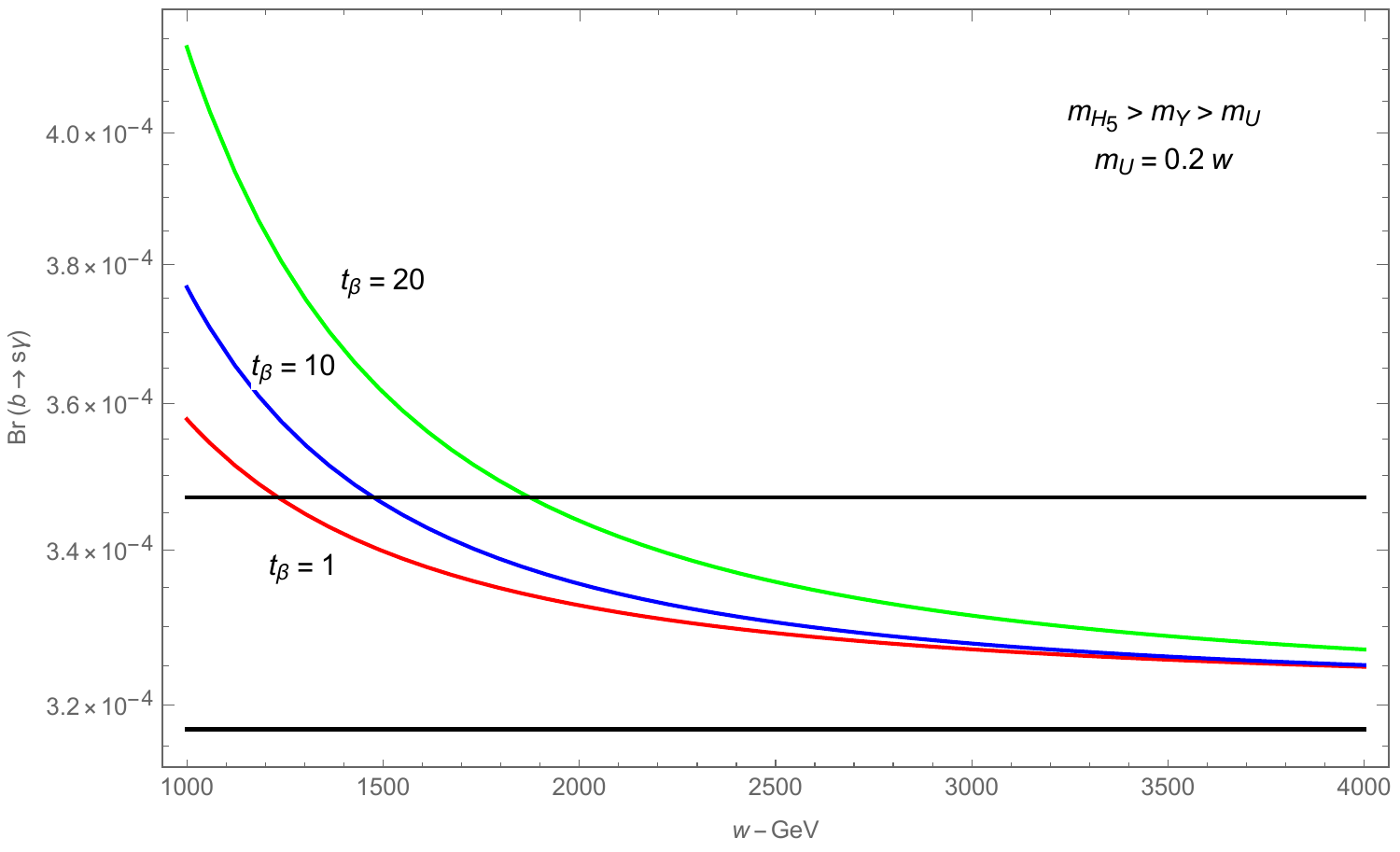}}
		\end{minipage}\hfill
		\begin{minipage}{0.5\textwidth}
			\includegraphics[width=\textwidth]{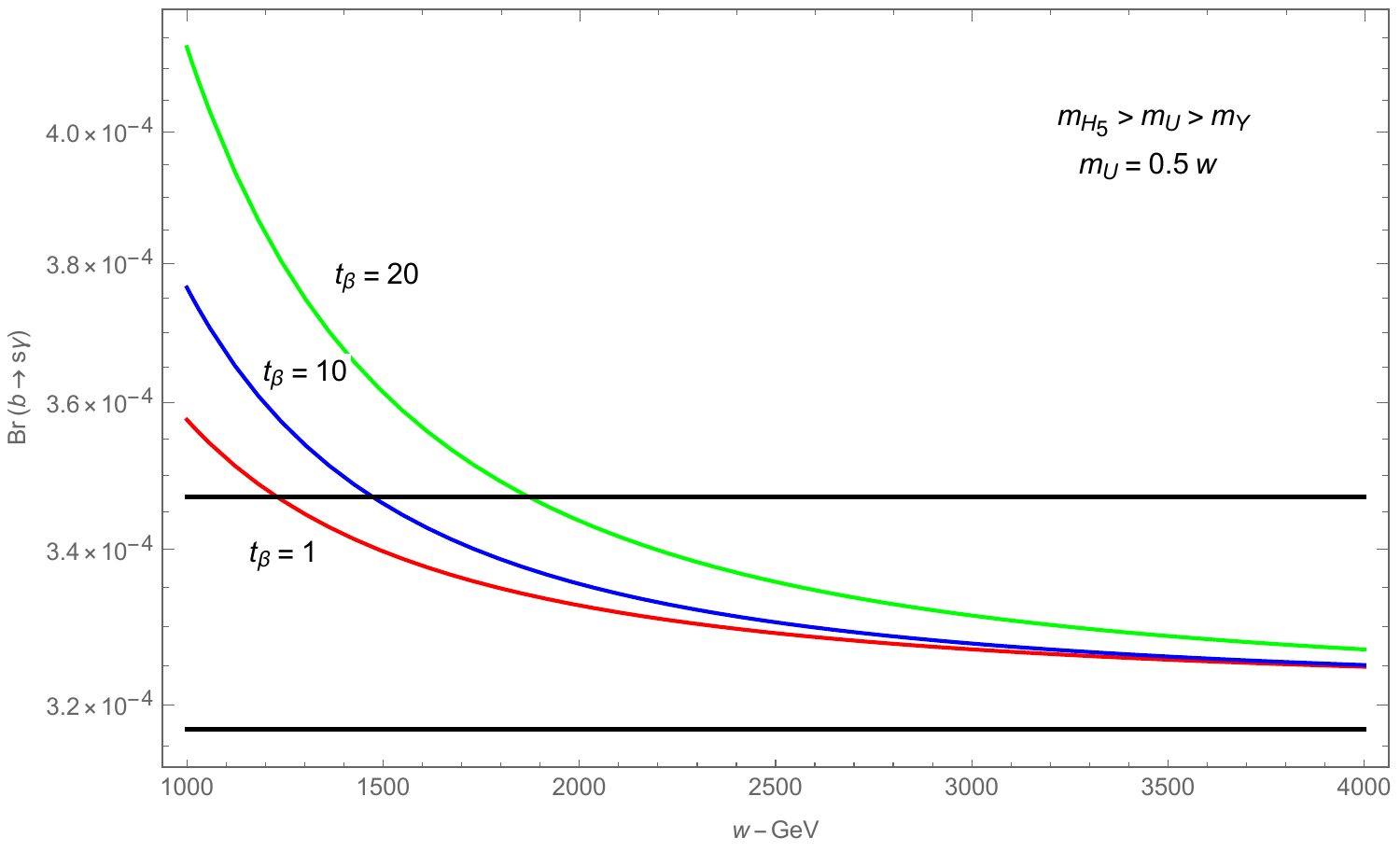}  
		\end{minipage}\par
		\includegraphics[width=0.5\textwidth]{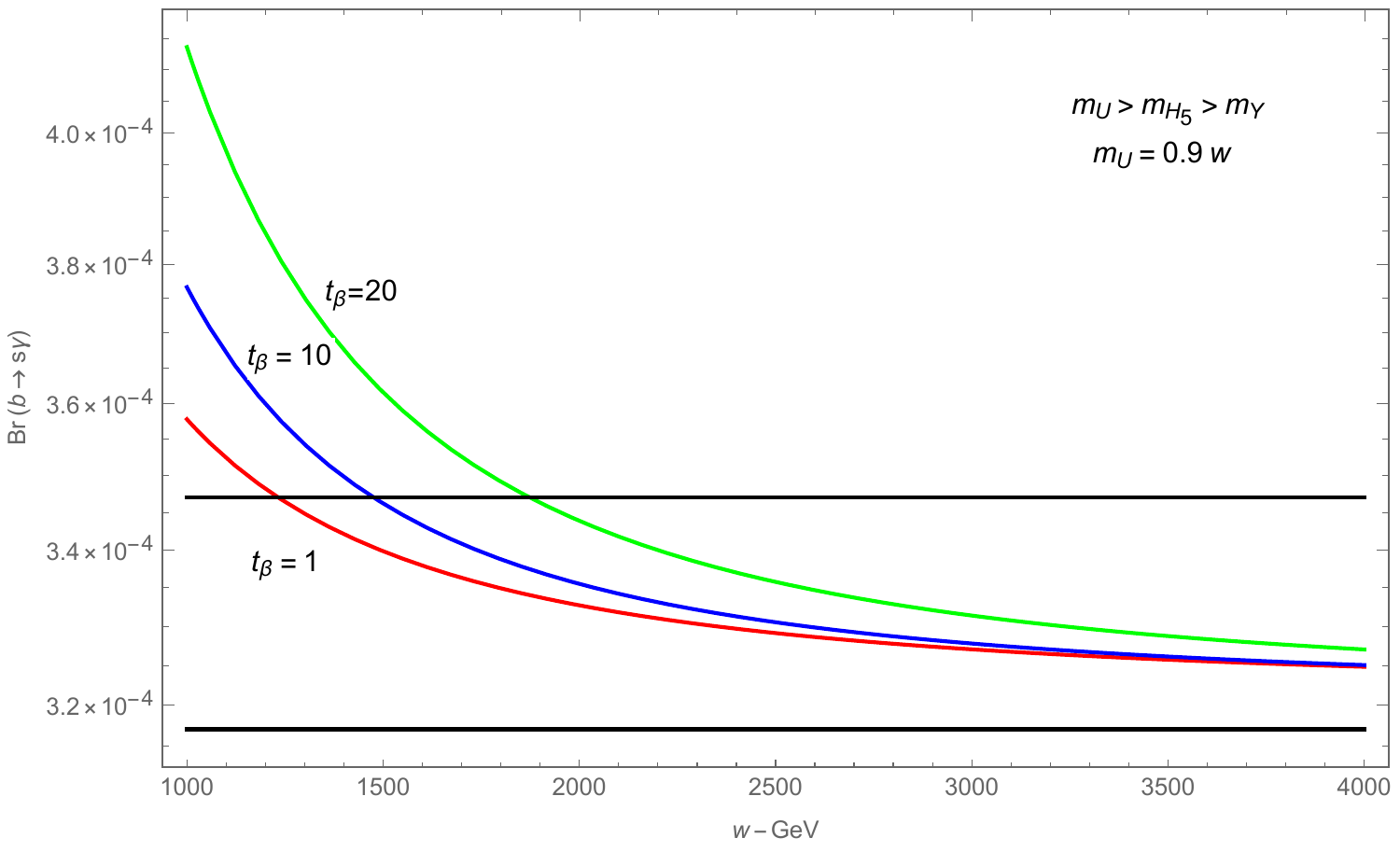}
		\caption{\label{plot-mY,H5-case II} The dependence of the branching ratio Br$(b \rightarrow s \gamma)$ on the \text{NP} scale $w$ in the limit $u,v \ll -f \sim w \sim \La$. The solid black lines indicate the current experimental constraint Br$(b\rightarrow s\gamma)=(3.32\pm0.15)\times 10^{-4}$ \cite{HFLAV}}.
		
	\end{figure}
	
	\subsection{Charged lepton flavor violation}
	The charged lepton flavor violation (CLFV) processes are strongly suppressed in the SM with right-handed neutrinos, Br$(l_i \rightarrow l_j \gamma) \simeq 10^{-55}$. Meanwhile, the current experimental bounds limits are given as 
	\cite{pdg}
	\bea
	&& \text{Br}(\mu^- \rightarrow e^- \gamma) <4.2 \times 10^{-13}, \crn
	&& \text{Br}(\tau^- \rightarrow e^- \gamma) <3.3 \times 10^{-8}, \crn
	&& \text{Br}(\tau^- \rightarrow \mu^- \gamma)< 4.4  \times 10^{-8}.
	\label{brm-eg}\eea
	It implies that the CLFV processes open a large window for studying the \text{NP} signals beyond the SM.
	Note that in the SM with right-handed neutrinos, the decay processes, $l_i \rightarrow l_j \gamma$, come from the one-loop level with $W^\pm$ mediated in the loop. The $\text{Br}(l_i \rightarrow l_j \gamma)$  is suppressed due to the mixing matrix elements of the neutrinos. The 3-3-1-1 model anticipates the existence of additional charged currents associated with the new charged particles, $Y^\pm, H^\pm_{4,5}$. Consequently, the new one-loop diagrams in the model may contribute significantly to the $\text{Br} ( l_i \rightarrow l_j \gamma)$. This branching ratio may reach the upper experimental bound given in Eq. (\ref{brm-eg}). In order to study the CLFV processes, we first write down the relevant Lagrangian based on the physical states as follows  
	\bea
	\mathcal{L}^{\text{lepton}}_{\mathrm{Scalar}}&&\supset \fr{h_{ab}^e u}{\sqrt{u^2+v^2}}\left(\bar{\nu}^{\prime}_{kL}(U^{\nu *}_L)_{ka}+\overline{(\nu^{\prime}_{kR})^c}
	V^{\nu *}_{ka}  \right)(U^l_{R})_{bj}e_{jR}^\prime H_5^+ +\fr{h_{ab}^e \omega}{\sqrt{v^2+\omega^2}}\overline{(N^\prime_{kR})^c}(U_R^{N })_{ka} (U^l_{ R})_{bj}e_{jR}^\prime H_4^+\nonumber \crn
	&&+\fr{h_{ab}^\nu v}{\sqrt{u^2+v^2}}\bar{e}^{\prime}_{jL}(U_{L}^{l*})_{ja}\left((V^{\nu T})_{bk}(\nu_{kL}^\prime)^c+(U^{\nu }_R)_{bk}\nu_{kR}^\prime \right)H_5^- \crn \nonumber && +\fr{h_{ab}^\nu \omega}{\sqrt{u^2+\omega^2}}\overline{(N^\prime_{jR})^c}(U_R^{N *})_{ja}\left((V^{\nu T})_{bk}(\nu_{kL}^\prime)^c+(U^{\nu }_R)_{bk}\nu_{kR}^\prime \right)H_o^\prime+ H.c.
	\label{Lmu-eg1} \eea   
	The charged currents associated with the new gauge bosons are written in the physical states as follows
	\bea
	\mathcal{L}_{W,Y}^{\text{lepton}}&&=-\fr{g}{\sqrt{2}} \left\{\overline{\nu}_{aL}\ga^\mu e_{aL}W_\mu^+ +\overline{e}_{aL}\ga^\mu (N_{aR})^cY_\mu^- \right\}+H.c. \crn && =-\fr{g}{\sqrt{2}}\left\{\left(\bar{\nu}^{\prime}_{kL}(U^{\nu *}_L)_{ka}+
	\overline{(\nu_{kR}^\prime)^c}V^{\nu *}_{ka} \right)\ga^\mu(U^l_{ L})_{aj}e_{jL}^\prime W^+_\mu + \bar{e}^{\prime}_{kL} (U^{l*}_L)_{ka}\ga^\mu (U_{R}^{N*})_{aj}(N^\prime_{jR})^c Y^-_\mu\right\}+H.c..
	\crn \label{Lmu-eg2}\eea
	Next, we write the effective Lagrangian relevant for the $\mu \rightarrow e \gamma$ processes in the traditional form 
	\bea
	\mathcal{L}_{\text{eff}}^{\mu \rightarrow e\gamma}= -4\fr{e G_F}{\sqrt{2}}m_\mu \left(A_R \bar{e} \sigma_{\mu \nu}P_R \mu +A_L \bar{e} \sigma_{\mu \nu}P_L \mu\right)F^{\mu \nu}+H.c.,
	\label{eff2}\eea
	where the factors $A_L, A_R$ are obtained by calculating all the one-loop diagrams. We use the 't Hooft-Feynman gauge and keep the external lepton masses for calculations. The obtained results are inspired by \cite{Lavoura}. The factors $A_{L, R}$ are divided into individual contributions, as shown below 
	\bea 
	A_{L,R} &=& A_{L,R}^W+A_{L,R}^Y+A_{L,R}^{H_5}+A_{L,R}^{H_4},\eea 
	where
	\bea 
	A_R^W &=& -\fr{eg^2}{32\pi^2 m_W^2} \sum_{j=1}^{3} (U^{\nu *}_L)_{\mu j} (U^{\nu }_L)_{ e j}f\left( \fr{m_{\nu_j}^2}{m_W^2}\right), \crn
	A_L^W &=& -\fr{eg^2 m_e}{32\pi^2 m_W^2 m_{\mu}} \sum_{j=1}^{3}  (U^{\nu *}_L)_{\mu j} (U^{\nu }_L)_{ e j} f\left(\fr{m_{\nu_j}^2}{m_W^2}\right), \crn
	A_R^Y &=& -\fr{eg^2}{32\pi^2 m_Y^2} \sum_{j=1}^{3} (U^{N *}_R)_{\mu j} (U^{N }_R)_{e j} f\left( \fr{m_{N_j}^2}{m_Y^2}\right), \crn
	A_L^Y &=& -\fr{eg^2 m_e}{32\pi^2 m_Y^2 m_{\mu}} \sum_{j=1}^{3}  (U^{N *}_R)_{\mu j} (U^{N}_R)_{ e j} f\left(\fr{m_{N_j}^2}{m_Y^2}\right), \nonumber \crn              
	A_L^{H_5} &=&-\fr{eg^2 m_e m_{\mu}}{32\pi^2 m_W^2 m_{H_5}^2t_{\beta}^2} \sum_{j=1}^{3} (U^{\nu *}_L)_{\mu j} (U^{\nu }_L)_{ e j} g\left( \fr{m_{\nu_j}^2}{m_{H_5}^2}\right) \crn &-& 
	\fr{eg^2  m_e v^2}{64\pi^2 m_W^2 m_{H_5}^2 m_{\mu}} \sum_{j,k=1}^{3} (h^{\nu *})_{\mu j}(h^{\nu})_{e j} (U^{\nu}_R)_{jk} (U^{\nu *}_R)_{j k}g\left( \fr{M_{\nu_{j}}^2}{m_{H_5}^2}\right) \crn &-& 
	\fr{eg^2   v^2m_e}{64\pi^2 m_W^2 m_{H_5}^2 m_{\mu}} \sum_{j,k=1}^{3} (h^{\nu *})_{\mu j}(h^{\nu})_{e j}(V^{\nu T})_{jk} (V^{\nu T*})_{j k}g\left( \fr{M_{\nu_{j}}^2}{m_{H_5}^2}\right), \crn 
	A_R^{H_5} &=&-\fr{eg^2 m_e^2}{32\pi^2 m_W^2 m_{H_5}^2 t_{\beta}^2} \sum_{j=1}^{3}  (U^{\nu *}_L)_{\mu j} (U^{\nu }_L)_{ e j} g\left( \fr{m_{\nu_j}^2}{m_{H_5}^2}\right) \crn &-& 
	\fr{eg^2  v^2}{64\pi^2 m_W^2 m_{H_5}^2 } \sum_{j,k=1}^{3} (h^{\nu *})_{\mu j}(h^{ \nu})_{e j}  (U^{\nu}_R)_{jk} (U^{\nu *}_R)_{j k} g\left( \fr{M_{\nu_{j}}^2}{m_{H_5}^2}\right)\crn &-& 
	\fr{eg^2   v^2}{64\pi^2 m_W^2 m_{H_5}^2 } \sum_{j,k=1}^{3} (h^{\nu *})_{\mu j}(h^{\nu})_{e j}(V^{\nu T})_{jk} (V^{\nu T*})_{j k}g\left( \fr{M_{\nu_{j}}^2}{m_{H_5}^2}\right), \crn
	A_L^{H_4} &=&-\fr{eg^2 m_e m_{\mu}}{32\pi^2 m_Y^2 m_{H_4}^2t_{\beta \prime }^2} \sum_{j=1}^{3} (U^{N *}_R)_{\mu j} (U^{N }_R)_{e j} g\left( \fr{m_{N_j}^2}{m_{H_4}^2}\right), \crn
	A_R^{H_4} &=&-\fr{eg^2 m_e^2}{32\pi^2 m_Y^2 m_{H_4}^2 t_{\beta \prime }^2} \sum_{j=1}^{3} (U^{N*}_R)_{\mu j} (U^{N }_R)_{e j} g\left( \fr{m_{N_j}^2}{m_{H_4}^2}\right),
	\eea 
	The functions $f(x)$ and $g(x)$ are defined by 
	\bea
	f(x) &=&\fr{10-43x+78x^2-48x^3+4x^4+18x^3\log{x}}{12 (x-1)^4},\crn
	g(x) &=& \fr{1-6x+3x^2+2x^3-6x^2\log{x}}{12(x-1)^4}. 
	\eea
	The notations $m_{\nu _j}, M_{\nu_{j}}, m_e, m_{\mu}$ are understood as the masses of light, heavy neutrinos, electron, and muon, respectively. 
	%The electromagnetic couplings $e, g$ are fixed by $ e=\sqrt{4\pi \al}, g=\fr{e}{s_W}$ with $\al =\fr{1}{137}$ is the %fine structure constant, $s_W$ is the Weinberg angle. All these parameters will be specified numerically latter. 
	From the effective Lagrangian (\ref{eff2}), we finally got the branching ratio $\text{Br}(\mu \rightarrow e \gamma)$ as follows
	\bea 
	\text{Br}(\mu \rightarrow e \gamma) &=& \fr{12\pi^2}{G_F^2}(|A_L|^2+|A_R|^2)\text{Br}(\mu \rightarrow e \tilde{\nu_{e}} \nu_{\mu}),
	\eea 
	where $G_F=\fr{g^2}{4\sqrt{2}m_W^2}$ is the Fermi coupling constant, $\text{Br}(\mu \rightarrow e \tilde{\nu_{e}} \nu_{\mu}) =100 \% $ as given in \cite{pdg}. 
	
	Before considering numerical calculations of the branching ratio $\text{Br}(\mu \rightarrow e \gamma )$, let us make some assumptions. We assume that a diagonal matrix presents the Yukawa couplings $h^e_{ab}$ in the flavor basis. 
	Thus, the matrix $U^{\nu}_L$ is identified as the PMNS matrix $U_{\text{PMNS}}$, which has been measured experimentally. Both the mixing matrices $U^\nu_R, V^{\nu}$ as well as $U_{L,R}^N$ are new and not constrained by experiments. To simplify, we suppose that the Yukawa couplings of the right-handed neutrinos $h^{ \prime \nu }$ are presented by a diagonal matrix. This indicates that the Majorana neutrino mass matrix has the form as $M_R^{\nu}=\text{Diag}(M_{\nu_1}, M_{\nu_2},M_{\nu_3})$ and thus the right-handed neutrino mixing mass matrix  $U^\nu_R$ is a unit matrix. The mixing matrix $V^{\nu}$ is also assumed to be diagonal. Finally, for the mixing matrix of the new leptons $U_{R}^N$, we can use three arbitrary angles $\theta^N_{ij}, (i,j=1,2,3)$ and a Dirac CP phase $\delta^N$ to parameterize. 

	With the above option, the Yukawa couplings $h^e, h^{\prime \nu}$ can be translated into the charged lepton and sterile neutrino masses as follows
	\bea
	h^e= -\frac{\sqrt{2}}{v} \text{Diag}\left (m_{e}, m_{\mu}, m_{\tau} \right), \hs h^{\prime \nu}=-\frac{1}{\sqrt{2}\Lambda}
	\text{Diag}\left(M_{\nu_1}, M_{\nu_2}, M_{\nu_3} \right).
	\eea  
	The Yukawa couplings $h^\nu$, which determine the neutrino Dirac mass, are rewritten by using Casas-Ibarra parametrization as given in \cite{JAcasas} 
	\bea h^{\nu}=\fr{\sqrt{2}}{u}\left(%
	\begin{array}{ccc}
		\sqrt{M_{\nu_1}} &0 &0\\
		0 & 	\sqrt{M_{ \nu_2}} & 0 \\
		0& 0& 	\sqrt{M_{\nu_3}} \\
	\end{array}%
	\right) R  \left(%
	\begin{array}{ccc}
		\sqrt{m_{\nu_1}} &0 & 0\\
		0 & 	\sqrt{m_{\nu_2}} & 0 \\
		0& 0& 	\sqrt{m_{\nu_3}} \\
	\end{array}%
	\right) U^{\nu \dagger}_L,
	\eea
	where $R$ is an orthogonal matrix which is presented via arbitrary angles as the following
	\bea
	R&=& \left(%
	\begin{array}{ccc}
		\hat{c}_2\hat{c}_3 &-\hat{c}_1 \hat{s}_3-\hat{s}_1\hat{s}_2\hat{c}_3 &\hat{s}_1\hat{s}_3-\hat{c}_1\hat{s}_2\hat{c}_3\\
		
		\hat{c}_2\hat{s}_3 &\hat{c}_1 \hat{c}_3-\hat{s}_1\hat{s}_2\hat{s}_3 &-\hat{s}_1\hat{c}_3-\hat{c}_1\hat{s}_2\hat{s}_3 \\
		
		\hat{s}_2 &\hat{s}_1 \hat{c}_2 &\hat{s}_1\hat{c}_2\\
	\end{array}%
	\right)
	\eea 
	with $\hat{s}_i=\sin \hat{\theta}_{i}$, $\hat{c}_i=\cos \hat{\theta}_{i}, i=1,2,3$ and $\hat{\theta}_{ij}\in [0,\pi/2]$. 
	
 For the magnitudes of relevant masses and the VEVs, we also work on the 
limits  $u,v \ll w\sim\La$, $u^2+v^2=246^2 \ \text{GeV}^2$. 
%Moreover, in the Sec. \ref{b-sgamma}, the mixing angle $t_{\beta}$ is required $t_{\beta}=v/u\ge 10$ in order to the new %physics scale in TeV scale, this conditions translates to the following upper bound $u<24.5$ GeV. 
To be consistent with the unitary bound \cite{unitarybound}, we need the constraint: $m_N < 16m_{Y}$. The masses of new charged Higgs $H_{4,5}^{\pm}$ and new gauge boson $Y^{\pm}$ are approximately taken as similar in the Sec. \ref{b-sgamma}. In keeping with constraints from dark matter studies in \cite{3311a}, the new fermion mass is at the \text{TeV} scale. The mixing angle $t_{\beta^{\prime}}$ can be expressed via the energy scales $u,w$ such as $t_{\beta^{\prime}}=\sqrt{246^2-u^2}/w$. Other known parameters are taken from \cite{pdg} as given

	\bea
	&& m_W=80.385\  \text{GeV}, \hs m_e=0.5109989461 \ \text{MeV}, \hs  m_{\mu}=105.6583745 \ \text{MeV} , \crn 
	&& \sin^2(\theta_{12})=0.307, \hs \sin^2(\theta_{23})=0.51, \hs \sin^2(\theta_{13})=0.021,\hs \al=\fr{1}{137},  \crn 
	&& \Delta m_{12}^2=7.53 \times 10^{-5} \ \text{eV}^2, \hs \Delta m_{23}^2=2.45  \times 10^{-3} \ \text{eV}^2, \eea
	where $\theta_{ij}$ are the mixing angles of the neutrino mixing matrix. 
	
	In addition, the branching ratio $\text{Br}(\mu \rightarrow e \gamma)$ also depends on the unknown parameters, such as  
	six mixing angles ($\hat{\theta}_{ij}$, $\theta_{ij}^N$), one CP phase $\delta^N$, 
	the masses of new particles $ m_N, M_{\nu_i}$. In the following, we are going to present the results
	of numerical calculations for the case where unknown parameters are chosen as  
	\bea && \theta^N_{12}=\pi/6, \hs \theta^N_{13}=\pi/3, \hs \theta^N_{23}=\pi/4, \hs \delta^N=0, \crn
	&& \hat{\theta}_{1}=\pi/3, \hs \hat{\theta}_{2}=\pi/4, \hs \hat{\theta}_{3}=\pi/6, \crn
	&& m_{\nu_1}=0.01 \ \text{eV} , \hs M_{\nu_1}=10^9 \ \text{GeV}, \hs M_{\nu_2}=M_{\nu_3}=10^3 M_{\nu_1}, \crn
	&&
	m_{N_1}=2000 \ \text{GeV}, \hs m_{N_2}=2200 \ \text{GeV}, \hs m_{N_3}=2400 \ \text{GeV}.  \eea 
	
\begin{figure}[H]
		\centering
		\begin{tabular}{cc}
			\includegraphics[width=10cm]{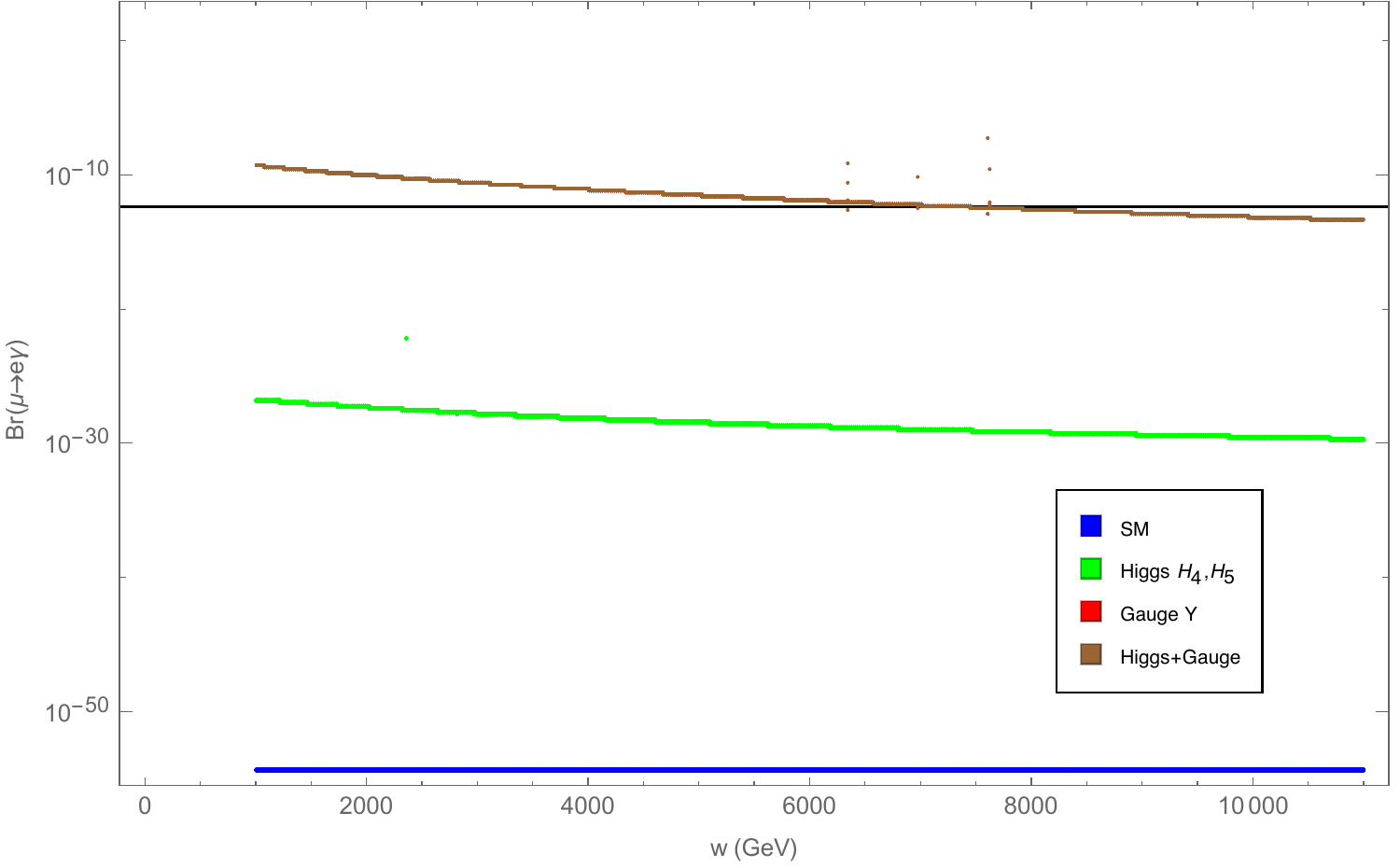}
		\end{tabular}
		\caption{\label{plot-mue}The figure presents the dependence of the branching ratio $\text{Br}(\mu \rightarrow e \gamma)$ on the NP scale $w$ for each contribution. The solid black line indicates the upper from the experiment \cite{pdg}. Here $u=10$ GeV.}
	\end{figure}

	The Fig. \ref{plot-mue} estimates the value of  each contribution into the  \text{Br}$(\mu \rightarrow e \gamma)$. The dominant contribution comes from the new gauge bosons $Y^\pm$. The \text{NP} scale is strongly constrained by the experiments \cite{pdg}, $\text{Br}(\mu \rightarrow e \gamma)_{\text{exp}} < 4.2 \times 10^{-13}$.  To be consistent with this bound,  the \text{NP} scale satisfies $w > 7.3$ \text{TeV}, which is similar to the bound derived from studying  the $b \rightarrow s \gamma$ decay.
	
\begin{figure}[H]
		\centering
		\begin{tabular}{cc}
			\includegraphics[width=10cm]{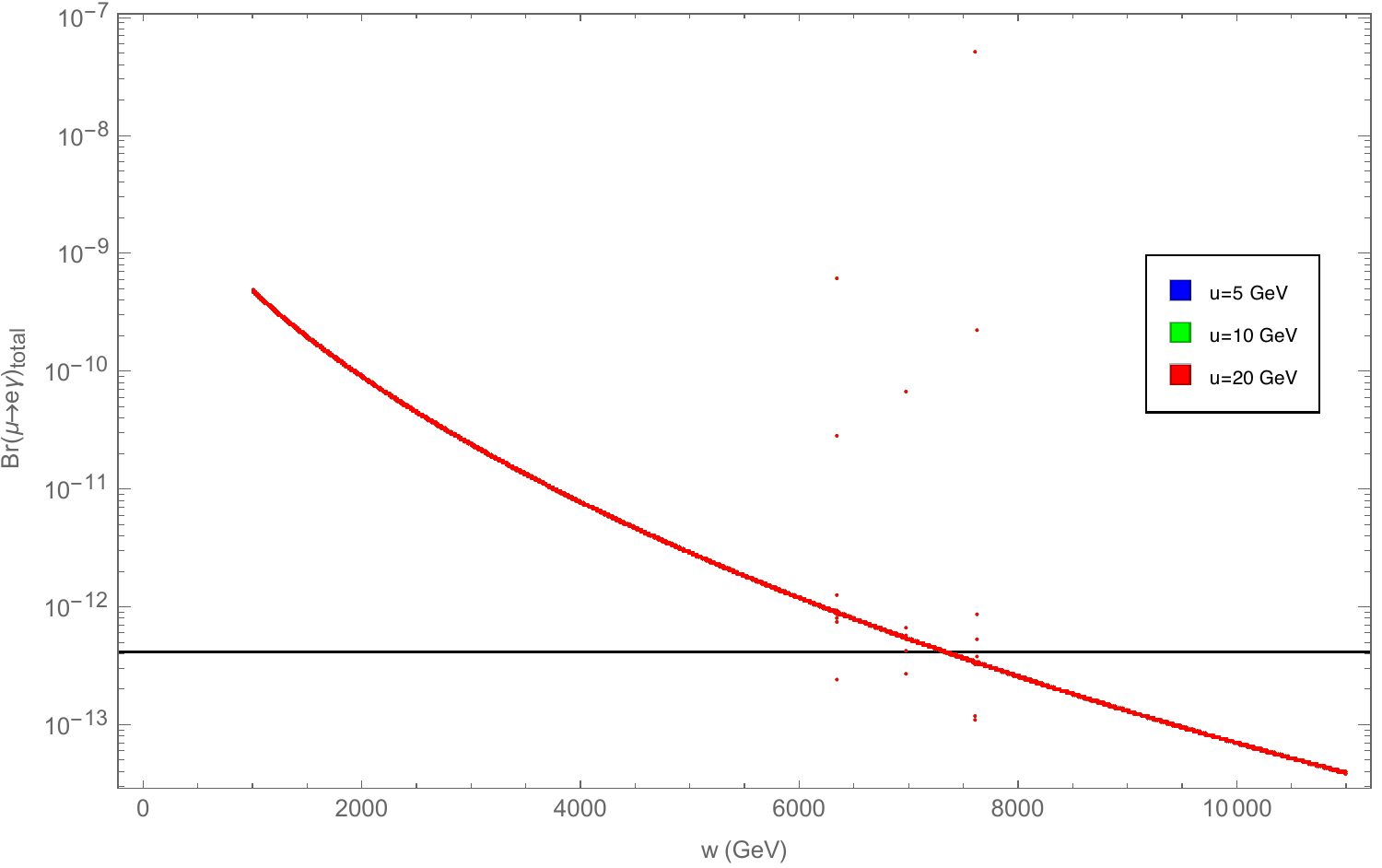}
		\end{tabular}
		\caption{\label{plot-total} The figure presents the comparison of the dependence of the total branching ratio Br$(\mu \rightarrow e \gamma)_{\text{total}}$ on the NP scale $w$ with $u=5$ GeV, $u=10$ GeV and $u=20$ GeV, respectively. The solid black line indicates the upper bound from the experiment \cite{pdg}.}
	\end{figure}

	The Fig. \ref{plot-total} demonstrates $\text{Br}(\mu \rightarrow e \gamma)_{\text{total}}$ as a function of \text{NP} scale $w$ with three different values of the electroweak scale, $u$, $u=5$ GeV, $u=10$ GeV and $u=20$ GeV. There is no separation between the graphs corresponding to different choices of $u$. As a result, the $\text{Br}(\mu \rightarrow e \gamma)_{\text{total}}$ depends very weakly on the $u$. It is important to keep in mind that the factors $A_{L,R}^{H_4,H_5}$ are greatly influenced by the electroweak scales $u$ and $v$. Therefore, this result shows that the charged currents associated with the charged Higgs particles have negligible influence on the $\mu \rightarrow e \gamma $ decay and may be ignored. Strong constraints are imposed on the charged current associated with new gauge bosons. 
	
	\section{\label{Conclusion}Conclusions}    
	In the 3-3-1-1 model, the tree-level FCNCs appear due to the non-universal assignment of quark families. Experiments on meson oscillations strongly constrain these interactions. We computed the mass difference for $ K^0 - \bar{K}^0, B_d^0- \bar{B}^0_d, B_s^0- \bar{B}^0_s $  based on the tree-level FCNCs and noticed that the main contributions to the meson oscillations come from the new neutral gauge bosons mediation. The NP scale is strongly constrained by the experimental bounds on mixing mass parameters. We have obtained the lower bound on the new gauge boson mass $\text{M}_{\text{new}}>12$ \text{TeV},  which is more stringent than the constraint previously given in \cite{3311a}. This change is because previous studies omitted the contributions of new Higgs, especially those of the SM. Our result is consistent with that of \cite{FCNCpond}. We also studied the tree-level FCNCs affecting the branching ratio of $B_s \rightarrow \mu^+ \mu^-$, $B\rightarrow K^* \mu^+ \mu^-$ and $B^{+}\rightarrow K^{+}\mu^{+}\mu^{-}$. In the parameter region consistent with the experimental constraints on the meson mass difference, the tree-level \text{FCNCs} give small contributions to these branching ratios, which is consistent with the measurement $B_s \rightarrow \mu^+ \mu^-$ \cite{Bsmm1,LHCb2,LHCb3,LHCb2021} but can not explain the $B \rightarrow K^* \mu^+ \mu^-$ and $B^{+}\rightarrow K^{+}\mu^{+}\mu^{-}$ anomalies \cite{AaiJ:2021,Aaij:2015dea,Khachatryan:2015isa,Wehle:2016yoi,Sirunyan:2017dhj,Aaboud:2018krd,Aaij:2020nrf, bsll2,bsll2-bs2}.
	
	For the radiative decay processes, we concentrated on the flavor-changing $b \rightarrow s\gamma$ decay. The large contribution arises from the Wilson coefficient $C_7^{H_5}$ yielded from one-loop diagrams with the new charged Higgs boson mediation. In spite of the enhanced contributions due to the factor $t_{\beta}=v/u$, the predicted branching ratio $\text{Br}(b \rightarrow s\gamma)$ is consistent with the measurement \cite{HFLAV}, if $\text{M}_{\text{new}}$ is chosen as above mentioned. In contrast to the $b \rightarrow s\gamma$ decay, the branching ratio of the lepton flavor-violating $\mu \rightarrow e \gamma$ decay obtains a large contribution from one-loop diagrams with new gauge bosons exchange. Due to the large mixing of new neutral leptons, the branching ratio $\text{Br}( \mu \rightarrow e \gamma)$ can reach the experimental upper bound.
	
	\section*{Acknowledgments}\vspace{-0.4cm}
	This research is funded by the Vietnam National Foundation for Science and Technology Development (NAFOSTED) 
	under Grant number 103.01-2019.312.
	
\end{document}